\documentclass[prd,aps,showpacs,epsf,floats,onecolumn]{revtex4}%
\usepackage{amssymb}
\usepackage{amsfonts}
\usepackage{amsmath}
\usepackage{graphicx}%
\providecommand{\U}[1]{\protect\rule{.1in}{.1in}}
\begin{document}
\title{\textbf{Aspects of Complexity in Quantum Evolutions on the Bloch Sphere}}
\author{\textbf{Carlo Cafaro}$^{1}$, \textbf{Emma Clements}$^{2}$, \textbf{Abeer
Alanazi}$^{2}$}
\affiliation{$^{1}$ Department of Nanoscale Science and Engineering, University at
Albany-SUNY, Albany, NY 12222, USA}
\affiliation{$^{2}$ Department of Physics, University at Albany-SUNY, Albany, NY 12222, USA}

\begin{abstract}
We enhance our quantitative comprehension of the complexity associated with
both time-optimal and time sub-optimal quantum Hamiltonian evolutions that
connect arbitrary source and target states on the Bloch sphere, as recently
presented in Nucl. Phys. \textbf{B1010, }116755 (2025). Initially, we examine
each unitary Schr\"{o}dinger quantum evolution selected through various
metrics, such as path length, geodesic efficiency, speed efficiency, and the
curvature coefficient of the corresponding quantum-mechanical trajectory that
connects the source state to the target state on the Bloch sphere.
Subsequently, we evaluate the selected evolutions using our proposed measure
of complexity, as well as in relation to the concept of complexity length
scale. The choice of both time-optimal and time sub-optimal evolutions, along
with the selection of source and target states, enables us to conduct
pertinent sanity checks that seek to validate the physical relevance of the
framework supporting our proposed complexity measure. Our research suggests
that, in general, efficient quantum evolutions possess a lower complexity than
their inefficient counterparts. However, it is important to recognize that
complexity is not solely determined by length; in fact, longer trajectories
that are adequately curved may exhibit a complexity that is less than or equal
to that of shorter trajectories with a lower curvature coefficient.

\end{abstract}

\maketitle

\section{Introduction}

Within the realm of quantum physics, there exist various concepts of
complexity. For example, one may refer to the complexity associated with a
quantum state (state complexity, \cite{chapman18,iaconis21}), the complexity
inherent in a quantum circuit (circuit complexity, \cite{nielsen,fernando21}),
or the complexity related to an operator (operator complexity,
\cite{parker19,roy23,liu23,caputa22}). Despite their distinct characteristics,
these measures of complexity share a common principle: the complexity of a
composite entity tends to increase in proportion to the number of fundamental
components required for its construction \cite{lof77,ay08,vijay22}. Depending
on the context, this relationship can often be articulated through
geometrically intuitive notions such as lengths and volumes. In the field of
theoretical computer science, Kolmogorov posited in Ref. \cite{kolmogorov68}
that the complexity of a sequence can be quantified by the length of the
shortest Turing machine program capable of generating it. Additionally, in the
domain of information theory, Rissanen suggested in Refs.
\cite{rissanen78,rissanen86} that the average minimal code length of a set of
messages serves as a measure of the complexity of that ensemble.

\medskip

The significance of the concepts of length and volume is crucial in defining
complexity within a quantum mechanical framework. For instance, state
complexity refers to the intricacy of a quantum state, articulated in terms of
the minimal local unitary circuit capable of producing the state from a basic
(i.e., factorizable) reference quantum state. Notably, a geometric
interpretation of state complexity was initially introduced during the
exploration of quantum state complexity in continuous many-body systems. In
particular, the state complexity of a target state, derived by applying a
series of parametrized unitary operators to a source state, was represented as
the length of the shortest path, as measured by the Fubini-Study metric,
corresponding to a permissible realization of the unitary operator
\cite{chapman18}. Quantum circuits consist of quantum gates that operate on
quantum states. Specifically, circuit complexity refers to the minimum number
of primitive gates required to construct a circuit that performs a
transformation on a specified quantum state \cite{nielsen,fernando21}. This
complexity is inherently discrete and serves as a valuable measure for
researchers engaged in the practical construction of quantum circuits from
fundamental gates. Notably, the concepts of state complexity and circuit
complexity are interconnected; state complexity pertains to the least complex
unitary operator that links the source and target states. The geometric
characterization of circuit complexity was introduced by Nielsen and his
colleagues in Refs. \cite{mike06,mike06B,mike08}. \ In this geometric
framework, the circuit complexity associated with a unitary operator $U$ is
inherently continuous and corresponds to the length of the shortest geodesic
connecting the identity operator to $U$ within the unitary group. The lengths
of these geodesic paths serve as a lower bound for the least number of quantum
gates required to construct the unitary operator $U$. Furthermore, operator
complexity pertains to the temporal increase in the size of an operator as it
evolves under the Heisenberg or Lindblad dynamics for closed and open quantum
systems, respectively. When assessed in relation to the Krylov basis, this
operator complexity is referred to as Krylov complexity \cite{parker19}. In
Ref. \cite{caputa22}, the temporal evolution of the displacement operator
during the unitary evolution of many-body quantum systems characterized by
symmetries revealed that the expectation value of the Krylov complexity
operator is equivalent to the volume of the associated classical phase space.
The relationship between Krylov complexity and volume was further explored
within specific quantum field theoretic contexts. Notably, the scaling of this
complexity with volume was substantiated in Ref. \cite{roy23} by showing that
Krylov complexity corresponds to the average particle number. This
relationship was effectively motivated by leveraging the proportionality
between volume and average particle number.

\medskip

This paper aims to extend the findings presented in Refs.
\cite{carloPRD,carloNPB} by emphasizing the essential roles of both length and
volume in understanding the complexity of quantum processes. The objective is
to deliver a comprehensive comparative analysis of the complexities associated
with both geodesic and non-geodesic quantum evolutions on the Bloch sphere for
two-level quantum systems. To deepen our understanding of the physical
implications of the newly introduced concept of complexity, this analysis will
also incorporate additional quantifiers that possess established physical
significance. These additional measures include path lengths, geodesic and
speed efficiencies, as well as curvature coefficients related to the quantum
evolutions. It is important to emphasize that although the concept of circuit
complexity is particularly relevant to digital quantum computing, which
involves discrete sequences of unitary logic gates, the complexity framework
presented in this paper is applicable to analog quantum computation,
characterized by continuous time evolutions.

\medskip

The main investigation carried out in this paper is justified by two specific
physically motivated questions for qubit systems.

\begin{enumerate}
\item[{[i]}] First, one expects that a proper complexity measure \textrm{C} of
quantum evolutions is such that
\begin{equation}
\mathrm{C}(\left\vert A\right\rangle \overset{\mathrm{H}_{\mathrm{opt}}%
}{\longrightarrow}\left\vert B\right\rangle )=\mathrm{C}(\left\vert A^{\prime
}\right\rangle \overset{\mathrm{H}_{\mathrm{opt}}^{\prime}}{\longrightarrow
}\left\vert B^{\prime}\right\rangle )\text{, if }s_{0}^{AB}=s_{0}^{A^{\prime
}B^{\prime}}\text{.} \label{love1}%
\end{equation}
In other words, is it true that the complexity of the evolution from
$\left\vert A\right\rangle $ to $\left\vert B\right\rangle $ under the time
optimal Hamiltonian $\mathrm{H}_{\mathrm{opt}}$ is equal to the complexity of
the evolution from $\left\vert A^{\prime}\right\rangle $ to $\left\vert
B^{\prime}\right\rangle $ under the time optimal Hamiltonian $\mathrm{H}%
_{\mathrm{opt}}^{\prime}$ provided that the distance $s_{0}^{AB}$ along the
shortest geodesic path that joins $\left\vert A\right\rangle $ to $\left\vert
B\right\rangle $ equals the distance $s_{0}^{A^{\prime}B^{\prime}}$ along the
shortest geodesic path that joins $\left\vert A^{\prime}\right\rangle $ to
$\left\vert B^{\prime}\right\rangle $?

\item[{[ii]}] Second, one expects that a proper complexity measure \textrm{C}
of quantum evolutions is such that
\begin{equation}
\mathrm{C}(\left\vert A\right\rangle \overset{\mathrm{H}_{\mathrm{sub}%
\text{-}\mathrm{opt}}}{\longrightarrow}\left\vert B\right\rangle
)=\mathrm{C}(\left\vert A^{\prime}\right\rangle \overset{\mathrm{H}%
_{\mathrm{sub}\text{-}\mathrm{opt}}^{\prime}}{\longrightarrow}\left\vert
B^{\prime}\right\rangle )\text{, if }s^{AB}=s^{A^{\prime}B^{\prime}}\text{.}
\label{love2}%
\end{equation}
In other words, is it true that the complexity of the evolution from
$\left\vert A\right\rangle $ to $\left\vert B\right\rangle $ under the time
sub-optimal Hamiltonian $\mathrm{H}_{\mathrm{sub}\text{-}\mathrm{opt}}$ is
equal to the complexity of the evolution from $\left\vert A^{\prime
}\right\rangle $ to $\left\vert B^{\prime}\right\rangle $ under the time
sub-optimal Hamiltonian $\mathrm{H}_{\mathrm{sub}\text{-}\mathrm{opt}}%
^{\prime}$ provided that the distance $s^{AB}$ along the effective dynamical
trajectory on the Bloch sphere that joins $\left\vert A\right\rangle $ to
$\left\vert B\right\rangle $ equals the distance $s^{A^{\prime}B^{\prime}}$
along the effective dynamical trajectory that joins $\left\vert A^{\prime
}\right\rangle $ to $\left\vert B^{\prime}\right\rangle $? When considering
Eq. (\ref{love2}), one also needs to assume that $s_{0}^{AB}=s_{0}^{A^{\prime
}B^{\prime}}$, with $s_{0}^{AB}\leq s^{AB}$ and $s_{0}^{A^{\prime}B^{\prime}%
}\leq s^{A^{\prime}B^{\prime}}$.
\end{enumerate}

It is crucial to address both of these questions for two primary reasons.
Firstly, it aids in grasping the importance of symmetry and invariance
arguments in relation to a complexity measure from a physical perspective.
Secondly, it facilitates a better understanding of the distinctions and
commonalities between the lengths and complexities of quantum evolutions,
whether they are optimal or sub-optimal in the context of travel time.

\medskip

The subsequent sections of this paper are structured as follows. In Section
II, we characterize quantum evolutions through the lens of path length
\cite{wootters81,provost80,cafaro23}, geodesic efficiency
\cite{anandan90,cafaro20}, speed efficiency \cite{uzdin12}, and curvature
coefficients \cite{alsing24A,alsing24B,cafaro25} associated with the dynamical
trajectories linking the specified initial and final states. In Section III,
we initiate our discussion within a classical probabilistic framework, where
the concept of information geometric complexity is typically employed to
assess the complexity of entropic motion on curved statistical manifolds,
particularly when information about the systems is limited
\cite{cafaro07,cafarothesis,cafaro17,cafaro18}. Transitioning to a
deterministic quantum framework, we introduce our definition of complexity in
quantum evolution, along with a concept of quantum complexity length scale. We
thoroughly examine the physical implications of these concepts, emphasizing
that both are articulated in terms of the accessed and accessible parametric
volumes of the regions on the Bloch sphere that delineate the evolution from
source to target states. In Section IV, we perform essential validation checks
aimed at confirming the physical significance of the framework underlying our
proposed complexity measure. These validations involve a direct comparison of
the numerical values associated with the complexities of appropriately
selected time-optimal and time sub-optimal evolutions derived from carefully
chosen source and target states. Specifically, we analyze three different
time-optimal evolutions, followed by an examination of two distinct time
sub-optimal evolutions. Each evolution is further assessed regarding its
geodesic efficiency, speed efficiency, and curvature coefficient. Finally, we
provide in Section V a summary of our findings along with concluding remarks.

\section{Quantifying Quantum Evolutions}

In this section, we begin with the presentation of the concepts of geodesic
efficiency, speed efficiency, and curvature coefficients associated with the
dynamical trajectories linking initial and final quantum states on the Bloch sphere.

\subsection{Geodesic efficiency}

The concept of geodesic efficiency $\eta_{\mathrm{GE}}$ for a quantum
evolution was at first proposed by Anandan and Aharonov in Ref.
\cite{anandan90}. The notion of geodesic efficiency provides insight into
quantum evolutions by measuring the deviations from the shortest path lengths
that link initial and final states along an unspecified trajectory.

Let us consider the evolution of a unit state vector $\left\vert \psi\left(
t\right)  \right\rangle $ that evolves according to the time-dependent
Schr\"{o}dinger equation, $i\hslash\partial_{t}\left\vert \psi\left(
t\right)  \right\rangle =\mathrm{H}\left(  t\right)  \left\vert \psi\left(
t\right)  \right\rangle $, with $t_{A}\leq t\leq t_{B}$. Then, the geodesic
efficiency $\eta_{\mathrm{GE}}$ that specifies such a quantum evolution over
the time interval $\left[  t_{A}\text{, }t_{B}\right]  $ is a scalar quantity
with $0\leq\eta_{\mathrm{GE}}\leq1$ given by \cite{anandan90,cafaro20}%
\begin{equation}
\eta_{\mathrm{GE}}\overset{\text{def}}{=}\frac{s_{0}}{s}=1-\frac{\Delta s}%
{s}=\frac{2\arccos\left[  \left\vert \left\langle A|B\right\rangle \right\vert
\right]  }{2\int_{t_{A}}^{t_{B}}\frac{\Delta E\left(  t\right)  }{\hslash}%
dt}\text{.} \label{efficiency}%
\end{equation}
In Eq. (\ref{efficiency}), $\Delta s\overset{\text{def}}{=}s-s_{0}$, $s_{0}$
represents the distance along the shortest geodesic path that joins the
initial $\left\vert A\right\rangle \overset{\text{def}}{=}$ $\left\vert
\psi\left(  t_{A}\right)  \right\rangle $ and final $\left\vert B\right\rangle
\overset{\text{def}}{=}\left\vert \psi\left(  t_{B}\right)  \right\rangle $
quantum states in the projective Hilbert space. The quantity $s$ in Eq.
(\ref{efficiency}), instead, denotes the distance along the dynamical
trajectory $\gamma\left(  t\right)  :t\mapsto\left\vert \psi\left(  t\right)
\right\rangle $ that corresponds to the quantum evolution of the unit state
vector $\left\vert \psi\left(  t\right)  \right\rangle $ with $t_{A}\leq t\leq
t_{B}$. Evidently, a geodesic quantum evolution with $\gamma\left(  t\right)
=\gamma_{\mathrm{geodesic}}\left(  t\right)  $ is characterized by the
condition $\eta_{\mathrm{GE}}^{(\gamma_{\mathrm{geodesic})}}=1$. Putting our
attention on the numerator in\ Eq. (\ref{efficiency}), we realize that it is
the angle between the unit state vectors $\left\vert A\right\rangle $ and
$\left\vert B\right\rangle $ and is equal to the Wootters distance
\cite{wootters81}. In particular, letting $\rho_{A}\overset{\text{def}}%
{=}\left\vert A\right\rangle \left\langle A\right\vert =(\mathbf{1+}\hat
{a}\cdot\mathbf{\boldsymbol{\sigma}})/2$ and $\rho_{B}\overset{\text{def}}%
{=}\left\vert B\right\rangle \left\langle B\right\vert =(\mathbf{1+}\hat
{b}\cdot\mathbf{\boldsymbol{\sigma}})/2$\textbf{ }with unit three-dimensional
vectors\textbf{ }$\hat{a}$ and $\hat{b}$ such that\textbf{ }$\hat{a}\cdot$
$\hat{b}=\cos(\theta_{AB})$,\textbf{ }it turns out that $s_{0}=\theta_{AB}$
given that $\left\vert \left\langle A|B\right\rangle \right\vert
^{2}=\mathrm{tr}\left(  \rho_{A}\rho_{B}\right)  +2\sqrt{\det(\rho_{A}%
)\det(\rho_{B})}=(1+\hat{a}\cdot\hat{b})/2=\cos^{2}\left(  \theta
_{AB}/2\right)  $. The quantity $\mathbf{\boldsymbol{\sigma}}\overset
{\text{def}}{=}\left(  \sigma_{x}\text{, }\sigma_{y}\text{, }\sigma
_{z}\right)  $ specifies the vector operator given in terms of the usual Pauli
operators $\sigma_{x}$, $\sigma_{y}$, and $\sigma_{z}$. Furthermore, the
denominator in Eq. (\ref{efficiency}) represents the integral of the
infinitesimal distance $ds\overset{\text{def}}{=}2\left[  \Delta E\left(
t\right)  /\hslash\right]  dt$ along the evolution curve in ray space (i.e.,
the projective Hilbert space) \cite{anandan90}. The quantity $\Delta E\left(
t\right)  \overset{\text{def}}{=}\left[  \left\langle \psi|\mathrm{H}%
^{2}\left(  t\right)  |\psi\right\rangle -\left\langle \psi|\mathrm{H}\left(
t\right)  |\psi\right\rangle ^{2}\right]  ^{1/2}$ describes the energy
uncertainty of the system in terms of the square root of the dispersion of the
Hamiltonian $\mathrm{H}\left(  t\right)  $.

It is known that Anandan and Aharonov demonstrated that the infinitesimal
distance $ds\overset{\text{def}}{=}2\left[  \Delta E\left(  t\right)
/\hslash\right]  dt$ is linked to the Fubini-Study infinitesimal distance
$ds_{\text{\textrm{FS}}}$ through the formula \cite{anandan90},%
\begin{equation}
ds_{\text{\textrm{FS}}}^{2}\left(  \left\vert \psi\left(  t\right)
\right\rangle \text{, }\left\vert \psi\left(  t+dt\right)  \right\rangle
\right)  \overset{\text{def}}{=}4\left[  1-\left\vert \left\langle \psi\left(
t\right)  |\psi\left(  t+dt\right)  \right\rangle \right\vert ^{2}\right]
=4\frac{\Delta E^{2}\left(  t\right)  }{\hslash^{2}}dt^{2}+\mathcal{O}\left(
dt^{3}\right)  \text{,} \label{relation}%
\end{equation}
where $\mathcal{O}\left(  dt^{3}\right)  $ is an infinitesimal quantity of an
order equal to or higher than $dt^{3}$. From the connection between
$ds_{\mathrm{FS}}$ and $ds$ in Eq. (\ref{relation}), one concludes that $s$ is
proportional to the time integral of $\Delta E$. In addition, $s$ specifies
the distance along the evolution of the quantum system in projective Hilbert
space as measured by the Fubini-Study metric. We remark that $\Delta s$ is
equal to zero and the geodesic efficiency $\eta_{\mathrm{GE}}$ in Eq.
(\ref{efficiency}) equals one when the actual dynamical trajectory
$\gamma\left(  t\right)  :t\mapsto\left\vert \psi\left(  t\right)
\right\rangle $ with $t_{A}\leq t\leq t_{B}$ is the shortest geodesic path
joining $\left\vert A\right\rangle $ and $\left\vert B\right\rangle $.
Naturally, the shortest possible distance $s_{0}$ between two orthogonal pure
states in ray space is equal to $\pi$. We point out that if we let
\textrm{H}$\left(  t\right)  \overset{\text{def}}{=}h_{0}\left(  t\right)
\mathbf{1+h}\left(  t\right)  \cdot\mathbf{\boldsymbol{\sigma}}$ and
$\rho\left(  t\right)  \overset{\text{def}}{=}(\mathbf{1+}\hat{a}\left(
t\right)  \cdot\mathbf{\boldsymbol{\sigma})/}2$ with $t_{A}\leq t\leq t_{B}$,
the energy uncertainty $\Delta E\left(  t\right)  \overset{\text{def}}{=}%
\sqrt{\mathrm{tr}\left(  \rho\mathrm{H}^{2}\right)  -\left[  \mathrm{tr}%
\left(  \rho\mathrm{H}\right)  \right]  ^{2}}$ becomes $\Delta E\left(
t\right)  =\sqrt{\mathbf{h}^{2}-\left[  \hat{a}\left(  t\right)
\cdot\mathbf{h}\right]  ^{2}}$. Lastly, the geodesic efficiency $\eta
_{\mathrm{GE}}$ in Eq. (\ref{efficiency}) can be rewritten as%
\begin{equation}
\eta_{\mathrm{GE}}=\frac{2\arccos\left(  \sqrt{\frac{1+\hat{a}\cdot\hat{b}}%
{2}}\right)  }{\int_{t_{A}}^{t_{B}}\frac{2}{\hslash}\sqrt{\mathbf{h}%
^{2}-\left[  \hat{a}\left(  t\right)  \cdot\mathbf{h}\right]  ^{2}}dt}\text{,}
\label{jap}%
\end{equation}
where $\hat{a}\left(  t_{A}\right)  \overset{\text{def}}{=}\hat{a}$ and
$\hat{a}\left(  t_{B}\right)  =\hat{b}$ in Eq. (\ref{jap}). Intriguingly, for
$\mathrm{H}\left(  t\right)  \overset{\text{def}}{=}\mathbf{h}\left(
t\right)  \cdot\mathbf{\boldsymbol{\sigma}}$ and letting $\mathbf{h}=\left[
\mathbf{h}\cdot\hat{a}\right]  \hat{a}+\left[  \mathbf{h}-(\mathbf{h}\cdot
\hat{a})\hat{a}\right]  =\mathbf{h}_{\shortparallel}+\mathbf{h}_{\perp}$, with
$\hat{a}=\hat{a}\left(  t\right)  $ in the decomposition of $\mathbf{h}$, the
geodesic efficiency $\eta_{\mathrm{GE}}$ in Eq. (\ref{jap}) can be recast as
\begin{equation}
\eta_{\mathrm{GE}}=\frac{\arccos\left(  \sqrt{\frac{1+\hat{a}\cdot\hat{b}}{2}%
}\right)  }{\int_{t_{A}}^{t_{B}}h_{\bot}(t)dt}\text{,} \label{goodyo1}%
\end{equation}
where we have set $\hslash=1$ in Eq. (\ref{goodyo1}). From Eq. (\ref{goodyo1}%
), one notices that $\eta_{\mathrm{GE}}$ is solely dependent on $h_{\bot}(t)$.
Retaining $\hslash=1$, it is worth noting that Feynman's geometric evolution
equation \cite{feynman57} $d\hat{a}/dt=2\mathbf{h}\times\hat{a}$ for the
time-dependent unit three-dimensional Bloch vector $\hat{a}=\hat{a}\left(
t\right)  $ can be regarded as a local (i.e., differential) formulation of the
Anandan-Aharonov relation $s=\int2\Delta E\left(  t\right)  dt$. As a matter
of fact, from $d\hat{a}/dt=2\mathbf{h}\times\hat{a}$, we obtain $da^{2}%
=4h_{\bot}^{2}dt^{2}$. Similarly, from $s=\int2\Delta E\left(  t\right)  dt$,
we get $ds=2h_{\bot}dt$. Therefore, exploiting these two differential
relations, we finally arrive at $da=2h_{\bot}dt=ds$.

We are now ready to present the concept of speed efficiency.

\subsection{Speed efficiency}

The notion of speed efficiency $\eta_{\mathrm{SE}}$ was originally proposed by
Uzdin and collaborators in Ref. \cite{uzdin12}. The concept of speed
efficiency, in a broad sense, aids in understanding quantum evolutions by
highlighting the extent to which energy resources are not utilized optimally.
It identifies the deviations from the minimal energy expenditure required to
facilitate the system's evolution along a specified trajectory.

More specifically, Uzdin and collaborators introduced in Ref. \cite{uzdin12}
classes of time-varying Hamiltonians appropriate for generating predetermined
dynamical trajectories specified by a minimal waste of energy resources.
Interestingly, despite being energetically resourceful, these dynamical
trajectories are usually not geodesics of minimum length. The requirement for
minimal waste of energy resources is specifically met when energy is not
expended on segments of the Hamiltonian that do not effectively guide the
system. In other words, all the available energy expressed in terms of the
spectral norm $\left\Vert \mathrm{H}\right\Vert _{\mathrm{SP}}$ of the
Hamiltonian $\mathrm{H}$ is converted into the speed of evolution of the
system $v_{\mathrm{H}}(t)$ $\overset{\text{def}}{=}(2/\hslash)\Delta E\left(
t\right)  $, where $\Delta E\left(  t\right)  $ represents the energy
uncertainty. In view of these considerations, one can be more explicit and
introduce Uzdin's speed efficiency $\eta_{\mathrm{SE}}$ with $0\leq
\eta_{\mathrm{SE}}\leq1$ . This is a a nonstationary (local) scalar quantity
defined as \cite{uzdin12}%
\begin{equation}
\eta_{\mathrm{SE}}\overset{\text{def}}{=}\frac{\Delta\mathrm{H}_{\rho}%
}{\left\Vert \mathrm{H}\right\Vert _{\mathrm{SP}}}=\frac{\sqrt{\mathrm{tr}%
\left(  \rho\mathrm{H}^{2}\right)  -\left[  \mathrm{tr}\left(  \rho
\mathrm{H}\right)  \right]  ^{2}}}{\max\left[  \sqrt{\mathrm{eig}\left(
\mathrm{H}^{\dagger}\mathrm{H}\right)  }\right]  }\text{.} \label{se1}%
\end{equation}
In Eq. (\ref{se1}), $\rho=\rho\left(  t\right)  \overset{\text{def}}%
{=}\left\vert \psi\left(  t\right)  \right\rangle \left\langle \psi\left(
t\right)  \right\vert $ denotes the density operator that characterizes the
quantum system at time $t$. Moreover, the quantity $\left\Vert \mathrm{H}%
\right\Vert _{\mathrm{SP}}$ located in the denominator of Eq. (\ref{se1}) is
given by $\left\Vert \mathrm{H}\right\Vert _{\mathrm{SP}}\overset{\text{def}%
}{=}\max\left[  \sqrt{\mathrm{eig}\left(  \mathrm{H}^{\dagger}\mathrm{H}%
\right)  }\right]  $ and represents the so-called spectral norm $\left\Vert
\mathrm{H}\right\Vert _{\mathrm{SP}}$ of the Hamiltonian operator \textrm{H.
}It measures the size of bounded linear operators $\left\{  \mathrm{H}%
\right\}  $ and is defined as the maximum of the square root of the
eigenvalues of the operator $\mathrm{H}^{\dagger}\mathrm{H}$, with
$\mathrm{H}^{\dagger}$ being the Hermitian conjugate of $\mathrm{H}$. Limiting
our attention to two-level quantum systems and employing the working condition
according to which the time-varying Hamiltonian $\mathrm{H}=\mathrm{H}\left(
t\right)  $ in Eq. (\ref{se1}) is expressed as $\mathrm{H}\left(  t\right)
\overset{\text{def}}{=}h_{0}\left(  t\right)  \mathbf{1}+\mathbf{h}\left(
t\right)  \cdot\mathbf{\boldsymbol{\sigma}}$, we note that the speed
efficiency $\eta_{\mathrm{SE}}$ in Eq. (\ref{se1}) can be suitably recast as%
\begin{equation}
\eta_{\mathrm{SE}}=\eta_{\mathrm{SE}}\left(  t\right)  \overset{\text{def}}%
{=}\frac{\sqrt{\mathbf{h}^{2}-(\hat{a}\cdot\mathbf{h})^{2}}}{\left\vert
h_{0}\right\vert +\sqrt{\mathbf{h}^{2}}}\text{.} \label{se2}%
\end{equation}
In Eq. (\ref{se2}), $\hat{a}\mathbf{=}\hat{a}\left(  t\right)  $ represents
the instantaneous three-dimensional unit Bloch vector that specifies the qubit
state of the system, while the set $\mathrm{eig}\left(  \mathrm{H}^{\dagger
}\mathrm{H}\right)  $ that enters the definition of $\left\Vert \mathrm{H}%
\right\Vert _{\mathrm{SP}}$ is equal to
\begin{equation}
\mathrm{eig}\left(  \mathrm{H}^{\dagger}\mathrm{H}\right)  \overset
{\text{def}}{=}\left\{  \lambda_{\mathrm{H}^{\dagger}\mathrm{H}}^{\left(
+\right)  }\overset{\text{def}}{=}(h_{0}+\sqrt{\mathbf{h}^{2}})^{2}\text{,
}\lambda_{\mathrm{H}^{\dagger}\mathrm{H}}^{\left(  -\right)  }\overset
{\text{def}}{=}(h_{0}-\sqrt{\mathbf{h}^{2}})^{2}\right\}  \text{.}%
\end{equation}
Given that the eigenvalues of $\mathrm{H}\left(  t\right)  \overset
{\text{def}}{=}h_{0}\left(  t\right)  \mathbf{1}+\mathbf{h}\left(  t\right)
\cdot\mathbf{\boldsymbol{\sigma}}$ are given by $E_{\pm}\overset{\text{def}%
}{=}h_{0}\pm\sqrt{\mathbf{h\cdot h}}$, the quantity $h_{0}=(E_{+}+E_{-})/2$
denotes the average of the two energy levels $E_{\pm}$ with $E_{+}\geq E_{-}$.
In addition, $\sqrt{\mathbf{h}^{2}}=(E_{+}-E_{-})/2$ is proportional to the
energy splitting $E_{+}-E_{-}$ between the two energy levels $E_{\pm}$.
Lastly, for a null trace nonstationary Hamiltonian $\mathrm{H}\left(
t\right)  \overset{\text{def}}{=}\mathbf{h}\left(  t\right)  \cdot
\mathbf{\boldsymbol{\sigma}}$ for which $\hat{a}(t)\cdot\mathbf{h}\left(
t\right)  =0$ \ at any time $t$, the speed efficiency $\eta_{\mathrm{SE}%
}\left(  t\right)  $ assumes its maximum value of one. Therefore, being
$\eta_{\mathrm{SE}}\left(  t\right)  =1$ for any $t$, the quantum evolution
occurs without any waste of energy resources. It is worth emphasizing that
when $\mathrm{H}\left(  t\right)  \overset{\text{def}}{=}\mathbf{h}\left(
t\right)  \cdot\mathbf{\boldsymbol{\sigma}}$, letting $\mathbf{h}%
=(\mathbf{h}\cdot\hat{a})\hat{a}+\left[  \mathbf{h}-(\mathbf{h}\cdot\hat
{a})\hat{a}\right]  =\mathbf{h}_{\shortparallel}+\mathbf{h}_{\perp}$ with
$\mathbf{h}_{\shortparallel}\cdot\mathbf{h}_{\perp}=0$, the speed efficiency
$\eta_{\mathrm{SE}}\left(  t\right)  $ in Eq. (\ref{se2}) can be fully recast
by means of the parallel (i.e., $\mathbf{h}_{\shortparallel}\overset
{\text{def}}{=}h_{\shortparallel}\hat{h}_{\shortparallel}$) and transverse
(i.e., $\mathbf{h}_{\perp}\overset{\text{def}}{=}h_{\perp}\hat{h}_{\perp}$)
components of the \textquotedblleft magnetic\textquotedblright\ field vector
$\mathbf{h}$,%
\begin{equation}
\eta_{\mathrm{SE}}\left(  t\right)  =\frac{h_{\bot}(t)}{\sqrt{h_{\bot}%
^{2}(t)+h_{\shortparallel}^{2}(t)}}\text{.} \label{eqtick}%
\end{equation}
It is transparent from Eq. (\ref{eqtick}) that $\eta_{\mathrm{SE}}\left(
t\right)  =1$ iff $h_{\shortparallel}(t)=0$ for any $t$ or, alternatively, iff
$\hat{a}\left(  t\right)  \cdot\mathbf{h}\left(  t\right)  =0$ for any $t$.
For completeness and focusing on qubit systems, we emphasize that the
Hamiltonians $\mathrm{H}\left(  t\right)  $ yielding\textbf{ }unit speed
efficiency trajectories in Ref. \cite{uzdin12} are constructed so that they
generate the same motion $\pi\left(  \left\vert \psi\left(  t\right)
\right\rangle \right)  $ in the complex projective Hilbert space $%
\mathbb{C}
P^{1}$ (or, alternatively, on the Bloch sphere $S^{2}\cong%
\mathbb{C}
P^{1}$) as $\left\vert \psi\left(  t\right)  \right\rangle $ (i.e.,
$\pi\left(  \left\vert m(t)\right\rangle \right)  =\pi\left(  \left\vert
\psi\left(  t\right)  \right\rangle \right)  $). Observe that the projection
operator $\pi$ can be formally specified as $\pi:\mathcal{H}_{2}^{1}%
\ni\left\vert \psi\left(  t\right)  \right\rangle \mapsto\pi\left(  \left\vert
\psi\left(  t\right)  \right\rangle \right)  \in%
\mathbb{C}
P^{1}$. Letting $\hslash=1$, it can be shown that an Hamiltonian
$\mathrm{H}\left(  t\right)  $ of this type can be formally recast as
\cite{uzdin12}%
\begin{equation}
\mathrm{H}\left(  t\right)  =i\left\vert \partial_{t}m(t)\right\rangle
\left\langle m(t)\right\vert -i\left\vert m(t)\right\rangle \left\langle
\partial_{t}m(t)\right\vert \text{,} \label{optH}%
\end{equation}
where, for simplicity of notation, we can set $\left\vert m(t)\right\rangle
=\left\vert m\right\rangle $ and $\left\vert \partial_{t}m(t)\right\rangle
=\left\vert \partial_{t}m\right\rangle =\partial_{t}\left\vert m\right\rangle
=\left\vert \dot{m}\right\rangle $. The state $\left\vert m\right\rangle $ in
Eq. (\ref{optH}) fulfills $\pi\left(  \left\vert m(t)\right\rangle \right)
=\pi\left(  \left\vert \psi\left(  t\right)  \right\rangle \right)  $,
$i\partial_{t}\left\vert m(t)\right\rangle =\mathrm{H}\left(  t\right)
\left\vert m(t)\right\rangle $, and $\eta_{\mathrm{SE}}\left(  t\right)  =1$.
The condition $\pi\left(  \left\vert m(t)\right\rangle \right)  =\pi\left(
\left\vert \psi\left(  t\right)  \right\rangle \right)  $ implies that
$\left\vert m(t)\right\rangle =c(t)\left\vert \psi\left(  t\right)
\right\rangle $, where $c(t)$ represents a complex scalar function. Imposing
that $\left\langle m\left\vert m\right.  \right\rangle =1$, we get $\left\vert
c(t)\right\vert =1$. For this reason, $c(t)$ can be expressed as
$e^{i\phi\left(  t\right)  }$ with $\phi\left(  t\right)  $ being a real
phase. At that point, introducing the parallel transport condition
$\left\langle m\left\vert \dot{m}\right.  \right\rangle =\left\langle \dot
{m}\left\vert m\right.  \right\rangle =0$, the phase $\phi\left(  t\right)  $
becomes $i\int\left\langle \psi\left\vert \dot{\psi}\right.  \right\rangle
dt$. Therefore, the state $\left\vert m(t)\right\rangle $ reduces to
$\left\vert m(t)\right\rangle =\exp(-\int_{0}^{t}\left\langle \psi(t^{\prime
})\left\vert \partial_{t^{\prime}}\psi(t^{\prime})\right.  \right\rangle
dt^{\prime})\left\vert \psi\left(  t\right)  \right\rangle $. Note that the
Hamiltonian $\mathrm{H}\left(  t\right)  $ in Eq. (\ref{optH}) is traceless
since its matrix representation with respect to the in the orthogonal basis
$\left\{  \left\vert m\right\rangle \text{, }\left\vert \partial
_{t}m\right\rangle \right\}  $ possesses just off-diagonal elements.
Furthermore, the relation $i\partial_{t}\left\vert m(t)\right\rangle
=\mathrm{H}\left(  t\right)  \left\vert m(t)\right\rangle $ implies that
$\left\vert m(t)\right\rangle $ satisfies the time-dependent Schr\"{o}dinger
evolution equation. Finally, the condition $\eta_{\mathrm{SE}}\left(
t\right)  =1$ signifies that $\mathrm{H}\left(  t\right)  $ evolves the state
$\left\vert m(t)\right\rangle $ with maximal speed and without any waste of
energy resources.

Having introduced the concepts of geodesic and speed efficiencies, we are now
ready to discuss the notion of curvature coefficient of a quantum evolution.

\subsection{Curvature}

The notion of curvature coefficient $\kappa_{\mathrm{AC}}^{2}$ of a quantum
evolution as employed in this paper was originally proposed by Alsing and
Cafaro in Refs. \cite{alsing24A,alsing24B}. As mentioned in Refs.
\cite{alsing24A,alsing24B}, the subscript \textquotedblleft\textrm{AC}%
\textquotedblright\textrm{\ }is an abbreviation of Alsing and Cafaro. The
curvature coefficient conceptually aids in comprehending quantum evolutions
from a geometric perspective, as it quantifies the extent to which a
trajectory linking two quantum states strays from a straight line within the
relevant geometric framework.

Although our focus in this paper is on two-level quantum systems, this
curvature concept can be formally specified for arbitrary finite-dimensional
quantum systems in a pure state. Let us consider a time-changing Hamiltonian
evolution that enters the Schr\"{o}dinger equation $i\hslash\partial
_{t}\left\vert \psi\left(  t\right)  \right\rangle =\mathrm{H}\left(
t\right)  \left\vert \psi\left(  t\right)  \right\rangle $. In general, the
unit quantum state $\left\vert \psi\left(  t\right)  \right\rangle $ is an
element of an arbitrary $N$-dimensional complex Hilbert space $\mathcal{H}%
_{N}$ with $N<+\infty$. In a typical scenario, $\left\vert \psi\left(
t\right)  \right\rangle $ is such that $\left\langle \psi\left(  t\right)
\left\vert \dot{\psi}\left(  t\right)  \right.  \right\rangle =(-i/\hslash
)\left\langle \psi\left(  t\right)  \left\vert \mathrm{H}\left(  t\right)
\right\vert \psi\left(  t\right)  \right\rangle \neq0$. However, given the
state $\left\vert \psi\left(  t\right)  \right\rangle $, we can construct the
parallel transported unit state vector $\left\vert \Psi\left(  t\right)
\right\rangle \overset{\text{def}}{=}e^{i\beta\left(  t\right)  }\left\vert
\psi\left(  t\right)  \right\rangle $ where the phase $\beta\left(  t\right)
$ is such that $\left\langle \Psi\left(  t\right)  \left\vert \dot{\Psi
}\left(  t\right)  \right.  \right\rangle =0$. Noting that $i\hslash\left\vert
\dot{\Psi}\left(  t\right)  \right\rangle =\left[  \mathrm{H}\left(  t\right)
-\hslash\dot{\beta}\left(  t\right)  \right]  \left\vert \Psi\left(  t\right)
\right\rangle $, the condition $\left\langle \Psi\left(  t\right)  \left\vert
\dot{\Psi}\left(  t\right)  \right.  \right\rangle =0$ is the same as setting
$\beta\left(  t\right)  $ equal to%
\begin{equation}
\beta\left(  t\right)  \overset{\text{def}}{=}\frac{1}{\hslash}\int_{0}%
^{t}\left\langle \psi\left(  t^{\prime}\right)  \left\vert \mathrm{H}\left(
t^{\prime}\right)  \right\vert \psi\left(  t^{\prime}\right)  \right\rangle
dt^{\prime}\text{.}%
\end{equation}
For this reason, the unit state vector $\left\vert \Psi\left(  t\right)
\right\rangle $ can be expressed as%
\begin{equation}
\left\vert \Psi\left(  t\right)  \right\rangle =e^{(i/\hslash)\int_{0}%
^{t}\left\langle \psi\left(  t^{\prime}\right)  \left\vert \mathrm{H}\left(
t^{\prime}\right)  \right\vert \psi\left(  t^{\prime}\right)  \right\rangle
dt^{\prime}}\left\vert \psi\left(  t\right)  \right\rangle \text{,}%
\end{equation}
and satisfies the evolution equation $i\hslash\left\vert \dot{\Psi}\left(
t\right)  \right\rangle =\Delta\mathrm{H}\left(  t\right)  \left\vert
\Psi\left(  t\right)  \right\rangle $ where $\Delta\mathrm{H}\left(  t\right)
\overset{\text{def}}{=}\mathrm{H}\left(  t\right)  -\left\langle
\mathrm{H}\left(  t\right)  \right\rangle $. As pointed out in Ref.
\cite{alsing24B}, the speed $v(t)$ of a quantum evolution is time-varying when
the Hamiltonian $\mathrm{H}=\mathrm{H}\left(  t\right)  $ changes in time. In
particular, $v(t)$ is such that $v^{2}\left(  t\right)  =\left\langle
\dot{\Psi}\left(  t\right)  \left\vert \dot{\Psi}\left(  t\right)  \right.
\right\rangle =\left\langle \left(  \Delta\mathrm{H}\left(  t\right)  \right)
^{2}\right\rangle /\hslash^{2}$. At this point, it is convenient to introduce
the arc length $s=s\left(  t\right)  $ specified by means of the speed
$v\left(  t\right)  $ as%
\begin{equation}
s\left(  t\right)  \overset{\text{def}}{=}\int_{0}^{t}v(t^{\prime})dt^{\prime
}\text{.} \label{s-equation}%
\end{equation}
From Eq. (\ref{s-equation}), we have $ds=v(t)dt$ and $\partial_{t}%
=v(t)\partial_{s}$. Then, making the introduction of the (adimensional)
operator $\Delta h\left(  t\right)  \overset{\text{def}}{=}\Delta
\mathrm{H}\left(  t\right)  /[\hslash v(t)]=\Delta\mathrm{H}\left(  t\right)
/\sqrt{\left\langle \left(  \Delta\mathrm{H}\left(  t\right)  \right)
^{2}\right\rangle }$, the normalized tangent vector $\left\vert T\left(
s\right)  \right\rangle \overset{\text{def}}{=}\partial_{s}\left\vert
\Psi\left(  s\right)  \right\rangle =\left\vert \Psi^{\prime}\left(  s\right)
\right\rangle $ can be expressed as $\left\vert T\left(  s\right)
\right\rangle =-i\Delta h\left(  s\right)  \left\vert \Psi\left(  s\right)
\right\rangle $. We observe that, by construction, $\left\langle T\left(
s\right)  \left\vert T\left(  s\right)  \right.  \right\rangle =1$. In
addition, we also have $\partial_{s}\left\langle \Delta h(s)\right\rangle
=\left\langle \Delta h^{\prime}(s)\right\rangle $. Intriguingly, this latter
equality remains valid\textbf{ }for arbitrary powers of differentiation. For
example, to the second power, we get $\partial_{s}^{2}\left\langle \Delta
h(s)\right\rangle =\left\langle \Delta h^{\prime\prime}(s)\right\rangle $. We
can construct $\left\vert T^{\prime}\left(  s\right)  \right\rangle
\overset{\text{def}}{=}\partial_{s}\left\vert T\left(  s\right)  \right\rangle
$ from the tangent vector $\left\vert T\left(  s\right)  \right\rangle
=-i\Delta h\left(  s\right)  \left\vert \Psi\left(  s\right)  \right\rangle $.
To be more explicit, we have $\left\vert T^{\prime}\left(  s\right)
\right\rangle =-i\Delta h(s)\left\vert \Psi^{\prime}\left(  s\right)
\right\rangle -i\Delta h^{\prime}(s)\left\vert \Psi\left(  s\right)
\right\rangle $ where, generally,
\begin{equation}
\left\langle T^{\prime}\left(  s\right)  \left\vert T^{\prime}\left(
s\right)  \right.  \right\rangle =\left\langle \left(  \Delta h^{\prime
}(s)\right)  ^{2}\right\rangle +\left\langle \left(  \Delta h(s)\right)
^{4}\right\rangle -2i\operatorname{Re}\left[  \left\langle \Delta h^{\prime
}(s)\left(  \Delta h(s)\right)  ^{2}\right\rangle \right]  \neq1\text{.}%
\end{equation}
Having introduced the vectors $\left\vert \Psi\left(  s\right)  \right\rangle
$, $\left\vert T\left(  s\right)  \right\rangle $, and $\left\vert T^{\prime
}\left(  s\right)  \right\rangle $, we can finally present the curvature
coefficient $\kappa_{\mathrm{AC}}^{2}\left(  s\right)  $ for quantum
evolutions that emerge from nonstationary Hamiltonians. The curvature
coefficient $\kappa_{\mathrm{AC}}^{2}\left(  s\right)  $ is defined as
$\kappa_{\mathrm{AC}}^{2}\left(  s\right)  \overset{\text{def}}{=}\left\langle
\tilde{N}_{\ast}\left(  s\right)  \left\vert \tilde{N}_{\ast}\left(  s\right)
\right.  \right\rangle $, where $\left\vert \tilde{N}_{\ast}\left(  s\right)
\right\rangle \overset{\text{def}}{=}\mathrm{P}^{\left(  \Psi\right)
}\left\vert T^{\prime}\left(  s\right)  \right\rangle $, $\mathrm{P}^{\left(
\Psi\right)  }\overset{\text{def}}{=}\mathrm{I}-\left\vert \Psi\left(
s\right)  \right\rangle \left\langle \Psi\left(  s\right)  \right\vert $, and
\textquotedblleft$\mathrm{I}$\textquotedblright\ denotes the identity operator
acting on states in $\mathcal{H}_{N}$. In terms of the notion of covariant
derivative $\mathrm{D}\overset{\text{def}}{=}\mathrm{P}^{\left(  \Psi\right)
}d/ds=\left(  \mathrm{I}-\left\vert \Psi\right\rangle \left\langle
\Psi\right\vert \right)  d/ds$ and $\mathrm{D}\left\vert T(s)\right\rangle
\overset{\text{def}}{=}\mathrm{P}^{\left(  \Psi\right)  }\left\vert T^{\prime
}(s)\right\rangle $ \cite{cafaro23,samuel88,paulPRA23}, the curvature
coefficient $\kappa_{\mathrm{AC}}^{2}\left(  s\right)  $ can be formally
recast as%
\begin{equation}
\kappa_{\mathrm{AC}}^{2}\left(  s\right)  \overset{\text{def}}{=}\left\Vert
\mathrm{D}\left\vert T(s)\right\rangle \right\Vert ^{2}=\left\Vert
\mathrm{D}^{2}\left\vert \Psi\left(  s\right)  \right\rangle \right\Vert
^{2}\text{.} \label{peggio}%
\end{equation}
From Eq. (\ref{peggio}), we note that $\kappa_{\mathrm{AC}}^{2}\left(
s\right)  $ is equal to the magnitude squared of the second covariant
derivative of the state vector $\left\vert \Psi\left(  s\right)  \right\rangle
$ employed to build the quantum-mechanical Schr\"{o}dinger trajectory in
projective Hilbert space. To be as transparent as possible, we stress that
$\left\vert \tilde{N}_{\ast}\left(  s\right)  \right\rangle \overset
{\text{def}}{=}\mathrm{P}^{\left(  \Psi\right)  }\left\vert T^{\prime}\left(
s\right)  \right\rangle $ is a vector that is not normalized to one and, in
addition, is not orthogonal to the vector $\left\vert T\left(  s\right)
\right\rangle $. Instead, although remaining unnormalized, $\left\vert
\tilde{N}\left(  s\right)  \right\rangle \overset{\text{def}}{=}%
\mathrm{P}^{\left(  T\right)  }\mathrm{P}^{\left(  \Psi\right)  }\left\vert
T^{\prime}\left(  s\right)  \right\rangle $ is orthogonal to $\left\vert
T\left(  s\right)  \right\rangle $. Ultimately, $\left\vert N\left(  s\right)
\right\rangle \overset{\text{def}}{=}$ $\left\vert \tilde{N}\left(  s\right)
\right\rangle /\sqrt{\left\langle \tilde{N}\left(  s\right)  \left\vert
\tilde{N}\left(  s\right)  \right.  \right\rangle }$ represents a normalized
vector which is orthogonal to $\left\vert T\left(  s\right)  \right\rangle $.
Summarizing, $\left\{  \left\vert \Psi\left(  s\right)  \right\rangle \text{,
}\left\vert T\left(  s\right)  \right\rangle \text{, }\left\vert N\left(
s\right)  \right\rangle \right\}  $ denotes the set of three orthonormal
vectors needed to specify the curvature coefficient $\kappa_{\mathrm{AC}}^{2}$
of a quantum evolution. We point out that, despite the fact that
$\mathcal{H}_{N}$ can be an arbitrary finite-dimensional complex space, we
focus our attention to the three-dimensional complex subspace spanned by the
set of vectors $\left\{  \left\vert \Psi\left(  s\right)  \right\rangle
\text{, }\left\vert T\left(  s\right)  \right\rangle \text{, }\left\vert
N\left(  s\right)  \right\rangle \right\}  $. Our working hypothesis aligns
with the traditional geometric viewpoint, which posits that the curvature and
torsion coefficients can be regarded as the first and second members,
respectively, of a broader family of generalized curvature functions
\cite{alvarez19}. In particular, for curves situated in higher-dimensional
spaces, this geometric viewpoint necessitates a collection of $m$ orthonormal
vectors to formulate $(m-1)$-generalized curvature functions \cite{alvarez19}.

The explicit calculation of the time-varying curvature coefficient
$\kappa_{\mathrm{AC}}^{2}\left(  s\right)  $ in Eq. (\ref{peggio}) via the
projection operators formalism is generally complicated. The complication is
justified by the fact that, similarly to what occurs in the classical
situation of space curves in $%
\mathbb{R}
^{3}$ \cite{parker77}, there are two main disadvantages when describing a
quantum curve by means of its arc length $s$. First of all, we may be not
capable of computing in closed form $s\left(  t\right)  $ in Eq.
(\ref{s-equation}). Second of all, even if we can arrive at an expression for
$s=s\left(  t\right)  $, we may be not capable of inverting this relation and,
thus, of arriving $t=t\left(  s\right)  $. The expression of $t=t\left(
s\right)  $, in turn, is necessary to find $\left\vert \Psi\left(  s\right)
\right\rangle \overset{\text{def}}{=}\left\vert \Psi\left(  t(s)\right)
\right\rangle $. To stay away from these difficulties, we can recast the
curvature coefficient $\kappa_{\mathrm{AC}}^{2}\left(  s\right)  $ in Eq.
(\ref{peggio}) in terms of expectation values evaluated with respect to the
state $\left\vert \Psi\left(  t\right)  \right\rangle $ (or, otherwise, with
respect to $\left\vert \psi\left(  t\right)  \right\rangle $). This state can
be calculated with no necessity of having the relation $t=t\left(  s\right)
$. For simplicity of notation, in what follows, we circumvent employing any
explicit reference to the $s$-dependence of the various operators and
expectation values being taken into consideration. For example, $\Delta
h\left(  s\right)  $ will be visible as $\Delta h$. After some algebraic
calculations, we arrive at%
\begin{equation}
\left\vert \tilde{N}_{\ast}\right\rangle =-\left\{  \left[  \left(  \Delta
h\right)  ^{2}-\left\langle \left(  \Delta h\right)  ^{2}\right\rangle
\right]  +i\left[  \Delta h^{\prime}-\left\langle \Delta h^{\prime
}\right\rangle \right]  \right\}  \left\vert \Psi\right\rangle \text{,}%
\label{nas}%
\end{equation}
where $\Delta h^{\prime}=\partial_{s}\left(  \Delta h\right)  =\left[
\partial_{t}\left(  \Delta h\right)  \right]  /v\left(  t\right)  $. For the
calculation of $\kappa_{\mathrm{AC}}^{2}\left(  s\right)  \overset{\text{def}%
}{=}\left\langle \tilde{N}_{\ast}\left(  s\right)  \left\vert \tilde{N}_{\ast
}\left(  s\right)  \right.  \right\rangle $, it is convenient to make use of
the Hermitian operator $\hat{\alpha}_{1}\overset{\text{def}}{=}\left(  \Delta
h\right)  ^{2}-\left\langle \left(  \Delta h\right)  ^{2}\right\rangle $ and
the anti-Hermitian operator $\hat{\beta}_{1}\overset{\text{def}}{=}i\left[
\Delta h^{\prime}-\left\langle \Delta h^{\prime}\right\rangle \right]  $ where
$\hat{\beta}_{1}^{\dagger}=-\hat{\beta}_{1}$. Then, $\left\vert \tilde
{N}_{\ast}\right\rangle =-\left(  \hat{\alpha}_{1}+\hat{\beta}_{1}\right)
\left\vert \Psi\right\rangle $ and $\left\langle \tilde{N}_{\ast}\left(
s\right)  \left\vert \tilde{N}_{\ast}\left(  s\right)  \right.  \right\rangle
$ is equal to $\left\langle \hat{\alpha}_{1}^{2}\right\rangle -\left\langle
\hat{\beta}_{1}^{2}\right\rangle +\left\langle \left[  \hat{\alpha}_{1}\text{,
}\hat{\beta}_{1}\right]  \right\rangle $ where $\left[  \hat{\alpha}%
_{1}\text{, }\hat{\beta}_{1}\right]  \overset{\text{def}}{=}\hat{\alpha}%
_{1}\hat{\beta}_{1}-\hat{\beta}_{1}\hat{\alpha}_{1}$ represents the quantum
commutator of $\hat{\alpha}_{1}$ and $\hat{\beta}_{1}$. Note that since
$\hat{\alpha}_{1}$ and $\hat{\beta}_{1}$ are Hermitian and anti-Hermitian
operators, respectively, $\left[  \hat{\alpha}_{1}\text{, }\hat{\beta}%
_{1}\right]  $ is a Hermitian operator and, consequently, the expectation
value $\left\langle \left[  \hat{\alpha}_{1}\text{, }\hat{\beta}_{1}\right]
\right\rangle $ is a real number. Making use of the definitions of
$\hat{\alpha}_{1}$ and $\hat{\beta}_{1}$, we arrive at $\left\langle
\hat{\alpha}_{1}^{2}\right\rangle =\left\langle (\Delta h)^{4}\right\rangle
-\left\langle (\Delta h)^{2}\right\rangle ^{2}$, $\left\langle \hat{\beta}%
_{1}^{2}\right\rangle =-\left[  \left\langle (\Delta h^{\prime})^{2}%
\right\rangle -\left\langle \Delta h^{\prime}\right\rangle ^{2}\right]  $, and
$\left\langle \left[  \hat{\alpha}_{1}\text{, }\hat{\beta}_{1}\right]
\right\rangle =i\left\langle \left[  (\Delta h)^{2}\text{, }\Delta h^{\prime
}\right]  \right\rangle $. Note that, since $\left[  (\Delta h)^{2}\text{,
}\Delta h^{\prime}\right]  $ is a anti-Hermitian operator, $\left\langle
\left[  (\Delta h)^{2}\text{, }\Delta h^{\prime}\right]  \right\rangle $ is a
purely imaginary scalar quantity. We also stress for completeness that
$\left[  (\Delta h)^{2}\text{, }\Delta h^{\prime}\right]  $ is typically not a
null operator. Indeed, we have $\left[  (\Delta h)^{2}\text{, }\Delta
h^{\prime}\right]  =\Delta h\left[  \Delta h\text{, }\Delta h^{\prime}\right]
+\left[  \Delta h\text{, }\Delta h^{\prime}\right]  \Delta h$ where $\left[
\Delta h\text{, }\Delta h^{\prime}\right]  =\left[  \mathrm{H}\text{,
}\mathrm{H}^{\prime}\right]  $. Then, concentrating on nonstationary qubit
Hamiltonians given by \textrm{H}$\left(  s\right)  \overset{\text{def}}%
{=}\mathbf{h}\left(  s\right)  \cdot\mathbf{\boldsymbol{\sigma}}$, the
commutator $\left[  \mathrm{H}\text{, }\mathrm{H}^{\prime}\right]
=2i(\mathbf{h}\times\mathbf{h}^{\prime})\cdot\mathbf{\boldsymbol{\sigma}}$ is
generally non vanishing since the vectors $\mathbf{h}$ and $\mathbf{h}%
^{\prime}$ are usually not collinear. As usual, $\mathbf{\boldsymbol{\sigma}}$
denotes the vector operator whose components are specified by the Pauli
operators $\sigma_{x}$, $\sigma_{y}$, and $\sigma_{z}$. Putting all together,
we finally arrive at a computationally convenient expression for the curvature
coefficient $\kappa_{\mathrm{AC}}^{2}\left(  s\right)  $ in Eq. (\ref{peggio})
in an arbitrary finite-dimensional nonstationary setting of quantum systems in
a pure state given by%
\begin{equation}
\kappa_{\mathrm{AC}}^{2}\left(  s\right)  =\left\langle (\Delta h)^{4}%
\right\rangle -\left\langle (\Delta h)^{2}\right\rangle ^{2}+\left[
\left\langle (\Delta h^{\prime})^{2}\right\rangle -\left\langle \Delta
h^{\prime}\right\rangle ^{2}\right]  +i\left\langle \left[  (\Delta
h)^{2}\text{, }\Delta h^{\prime}\right]  \right\rangle \text{.}%
\label{curvatime}%
\end{equation}
It is worth pointing out that inspection of Eq. (\ref{curvatime})\textbf{
}leads to the realization that when the Hamiltonian \textrm{H} is not
time-varying, $\Delta h^{\prime}$ becomes to the null operator and we recover
the stationary limit $\left\langle (\Delta h)^{4}\right\rangle -\left\langle
(\Delta h)^{2}\right\rangle ^{2}$ of $\kappa_{\mathrm{AC}}^{2}\left(
s\right)  $ as reported in \ Ref. \cite{alsing24A}. The expression of the
curvature coefficient $\kappa_{\mathrm{AC}}^{2}\left(  s\right)  $ in Eq.
(\ref{curvatime}) manifests itself in terms of an approach that is based upon
the computation of expectation values which, in turn, necessitate of the
knowledge of \ the state vector $\left\vert \psi\left(  t\right)
\right\rangle $ that evolves according to the time-dependent Schr\"{o}dinger
evolution equation. As emphasized in Ref. \cite{alsing24B}, the
expectation-values approach offers an intuitive statistical meaning for
$\kappa_{\mathrm{AC}}^{2}\left(  s\right)  $. Nevertheless, this method lacks
a clear geometric significance. Due to this deficiency and by focusing
exclusively on time-varying Hamiltonians and two-level quantum systems, one
can derive a closed-form expression for the curvature coefficient associated
with a curve delineated by a single-qubit quantum state that evolves under the
influence of a general nonstationary Hamiltonian. Indeed, the curvature
coefficient $\kappa_{\mathrm{AC}}^{2}$ can be fully expressed by means of just
two real three-dimensional vectors with a neat geometrical significance.
Specifically, the two vectors are the Bloch vector $\mathbf{a}\left(
t\right)  $ and the magnetic field vector $\mathbf{h}\left(  t\right)  $.
While the former vector specifies the density operator $\rho\left(  t\right)
=$ $\left\vert \psi\left(  t\right)  \right\rangle \left\langle \psi\left(
t\right)  \right\vert \overset{\text{def}}{=}\left[  \mathrm{I}+\mathbf{a}%
\left(  t\right)  \cdot\mathbf{\boldsymbol{\sigma}}\right]  /2$, the latter
defines the nonstationary (traceless) Hamiltonian \textrm{H}$\left(  t\right)
\overset{\text{def}}{=}\mathbf{h}\left(  t\right)  \cdot
\mathbf{\boldsymbol{\sigma}}$. Following the unabridged derivation discussed
in Ref. \cite{alsing24B}, we arrive at%
\begin{equation}
\kappa_{\mathrm{AC}}^{2}\left(  \mathbf{a}\text{, }\mathbf{h}\right)
=4\frac{\left(  \mathbf{a\cdot h}\right)  ^{2}}{\mathbf{h}^{2}-\left(
\mathbf{a\cdot h}\right)  ^{2}}+\frac{\left[  \mathbf{h}^{2}\mathbf{\dot{h}%
}^{2}-\left(  \mathbf{h\cdot\dot{h}}\right)  ^{2}\right]  -\left[  \left(
\mathbf{a\cdot\dot{h}}\right)  \mathbf{h-}\left(  \mathbf{a\cdot h}\right)
\mathbf{\dot{h}}\right]  ^{2}}{\left[  \mathbf{h}^{2}-\left(  \mathbf{a\cdot
h}\right)  ^{2}\right]  ^{3}}+4\frac{\left(  \mathbf{a\cdot h}\right)  \left[
\mathbf{a\cdot}\left(  \mathbf{h\times\dot{h}}\right)  \right]  }{\left[
\mathbf{h}^{2}-\left(  \mathbf{a\cdot h}\right)  ^{2}\right]  ^{2}}%
\text{.}\label{XXX}%
\end{equation}
The formulation of $\kappa_{\mathrm{AC}}^{2}$ as reported in Eq. (\ref{XXX})
provides a significant resource for computational investigations within qubit
systems. Moreover, it possesses a clear geometric interpretation of the
curvature of a quantum evolution in terms of the normalized unitless Bloch
vector $\mathbf{a}$ and, in addition, the typically unnormalized magnetic
field vector $\mathbf{h}$ with $\left[  \mathbf{h}\right]  _{\mathrm{MKSA}}%
=$\textrm{joules}$=\sec.^{-1}$ when setting $\hslash=1$. Finally, we remark
that for stationary Hamiltonians with non time-varying $\mathbf{h}$,
$\kappa_{\mathrm{AC}}^{2}\left(  \mathbf{a}\text{, }\mathbf{h}\right)  $ in
Eq. (\ref{XXX}) reduces to%
\begin{equation}
\kappa_{\mathrm{AC}}^{2}\left(  \mathbf{a}\text{, }\mathbf{h}\right)
=4\frac{\left(  \mathbf{a\cdot h}\right)  ^{2}}{\mathbf{h}^{2}-\left(
\mathbf{a\cdot h}\right)  ^{2}}\text{.}\label{XXXX}%
\end{equation}
Clearly, the notion of curvature requires the introduction of the concept of
parallel transport. This, in turn, is described by means of a connection and a
covariant derivative. For an explicit expression of the Christoffel symbols of
the Levi-Civita connection on the Bloch sphere equipped with the Fubini-Study
metric and patched with a set of affine coordinates, we suggest Ref.
\cite{karol06}. In our examination, we employ the covariant derivative
$\mathrm{D}\overset{\text{def}}{=}\mathrm{P}^{\left(  \Psi\right)  }d/ds$ as
the differential operator for establishing and utilizing the Levi-Civita
connection on the Bloch sphere.

Before moving to the next section, we stress that it is noteworthy that
time-optimal quantum evolutions can be identified through variational methods,
similar to the approach taken in classical mechanics when addressing the
brachistochrone problem \cite{carlini06}. In particular, the qubit geodesics
on the Bloch sphere can be ascertained by minimizing the integral $\int
ds_{\mathrm{FS}}$\textbf{ }across all paths that link pairs of points on the
Bloch sphere, where $ds_{\mathrm{FS}}$ represents the Fubini-Study
infinitesimal line element \cite{erik20}. More generally, geometric concepts
relevant to the examination of quantum state manifolds encompass the idea of
quantum Christoffel symbols, as detailed in Refs. \cite{popov23,hetenyi23}.
Specifically, as outlined in Ref. \cite{hetenyi23}, quantum geometric
quantities can be linked to quantum fluctuations. For example, the second
cumulant serves to generate the metric tensor, while the third cumulant is
associated with the Christoffel symbol, and the fourth cumulant relates to the
Riemann curvature tensor. Indeed, focusing on the relationship between
complexity and curvature, we believe that conducting a comparative analysis
between our curvature coefficient\textbf{ }$\kappa_{\mathrm{AC}}^{2}$ and the
four-index Riemannian curvature presented in Ref. \cite{hetenyi23} would be
highly beneficial. The four-index Riemannian curvature discussed in Ref.
\cite{hetenyi23} has been demonstrated to correspond directly to the fourth
cumulant, or kurtosis, of a cumulant generating function that describes the
Riemannian geometry of the quantum system's state space. In our approach
\cite{alsing24A,alsing24B}, the curvature coefficient is linked to an
expression for fourth-order quantum-mechanical fluctuations, which, when
viewed from a statistical standpoint, relates to the notion of kurtosis. We
reserve this fascinating scientific comparison for future investigations.

Having presented the notions of geodesic and speed efficiencies $\eta
_{\mathrm{GE}}$ and $\eta_{\mathrm{SE}}$ in Eqs. (\ref{jap}) and (\ref{se2}),
respectively, together with the concept of curvature coefficient
$\kappa_{\mathrm{AC}}^{2}$ of a quantum evolution in Eq. (\ref{XXX}), we
discuss our proposed concept of complexity of a quantum evolution in the next section.

\section{Complexity}

The concepts of complexity of a quantum evolution together with the notion of
complexity length scale as employed in this paper were originally introduced
by one of the authors in Ref. \cite{carloNPB}.

To begin, we stress that our proposal for a suitable quantum-mechanical notion
of complexity originates from a classical measure of complexity developed in
the context of a probabilistic description of physical systems in the presence
of partial information and limited knowledge
\cite{cafaro07,cafarothesis,ali18,ali21,cafaro17,cafaro18}. In what follows,
we aim to be as comprehensive as possible. Nevertheless, we recommend
consulting the original presentation of our proposed complexity measure as
detailed in Ref. \cite{carloNPB}. In the shift from a probabilistic classical
framework to a deterministic quantum framework concerning two-level quantum
systems, several distinctions arise. Initially, we move from a geodesic motion
driven by entropic arguments within an arbitrary $N_{\mathrm{c}}$-dimensional
curved statistical manifold of probability distributions to a deterministic
quantum temporal evolution, which may be either geodesic or non-geodesic, on a
two-dimensional Bloch sphere. Any unit pure quantum state that evolves
dynamically on this sphere can be described using two real parameters, such as
the polar and azimuthal angles. Secondly, while the Fisher-Rao information
metric serves as the natural metric for classical statistical manifolds, the
Fubini-Study metric is the appropriate metric for the Bloch sphere. Thirdly,
the parameter space that is globally accessible in the quantum realm is
finite, contrasting with the classical probabilistic domain, where, for
example, the parameter space of Gaussian probability distributions is not
finite. Fourthly, we anticipate that the intricate details in quantum
evolution will be considerably less complex than those observed in classical
probabilistic evolution. Nonetheless, the technique of temporal averaging
remains significant in the deterministic quantum context discussed here, as it
helps to smooth out finer details and focus on the more relevant overall
global behavior of quantum evolution. Fifthly, the entropic nature of
classical evolution necessitates an examination of the asymptotic temporal
behavior of statistical volumes to eliminate temporary transient effects and
concentrate solely on enduring effects. However, due to the deterministic
characteristics of the quantum evolution under consideration, this long-time
limit is unnecessary, as there is no need to distinguish between
(non-permanent) short-term and (permanent) long-term behaviors. Furthermore,
we are examining finite-time quantum processes, for which the notion of
asymptotic temporal behavior is not properly defined.

In light of these initial comments, in order to measure the intricate dynamics
of quantum changes within a specific time span $\left[  t_{A}\text{, }%
t_{B}\right]  $, we introduce the value \textrm{C}$\left(  t_{A}\text{, }%
t_{B}\right)  $ given by%
\begin{equation}
\mathrm{C}\left(  t_{A}\text{, }t_{B}\right)  \overset{\text{def}}{=}%
\frac{\mathrm{V}_{\max}\left(  t_{A}\text{, }t_{B}\right)  -\overline
{\mathrm{V}}\left(  t_{A}\text{, }t_{B}\right)  }{\mathrm{V}_{\max}\left(
t_{A}\text{, }t_{B}\right)  }\text{.} \label{QCD}%
\end{equation}
The physical justification for proposing this expression for the complexity
$\mathrm{C}\left(  t_{A}\text{, }t_{B}\right)  $ will be explained in what
follows. We begin by defining $\overline{\mathrm{V}}\left(  t_{A}\text{,
}t_{B}\right)  $ and $\mathrm{V}_{\max}\left(  t_{A}\text{, }t_{B}\right)  $
in Eq. (\ref{QCD}). To present the definition of the so-called \emph{accessed
volume} $\overline{\mathrm{V}}\left(  t_{A}\text{, }t_{B}\right)  $, we
proceed in a schematic fashion in the following manner. If possible integrate
analytically the time-dependent Schr\"{o}dinger evolution equation
$i\hslash\partial_{t}\left\vert \psi(t)\right\rangle =\mathrm{H}\left(
t\right)  \left\vert \psi(t)\right\rangle $ and express the (normalized)
single-qubit state vector $\left\vert \psi(t)\right\rangle $ at an arbitrary
time $t$ in terms of the computational basis state vectors $\left\{
\left\vert 0\right\rangle \text{, }\left\vert 1\right\rangle \right\}  $. If
not possible, proceed with a numerical analysis. We find $\left\vert
\psi(t)\right\rangle =c_{0}(t)\left\vert 0\right\rangle +c_{1}(t)\left\vert
1\right\rangle $ where $c_{0}(t)$ and $c_{1}(t)$ can be expressed as
\begin{equation}
c_{0}(t)\overset{\text{def}}{=}\left\langle 0\left\vert \psi(t)\right.
\right\rangle =\left\vert c_{0}(t)\right\vert e^{i\phi_{0}(t)}\text{, and
}c_{1}(t)\overset{\text{def}}{=}\left\langle 1\left\vert \psi(t)\right.
\right\rangle =\left\vert c_{1}(t)\right\vert e^{i\phi_{1}(t)}\text{,}
\label{qa}%
\end{equation}
respectively. Moreover, observe that $\phi_{0}(t)$ and $\phi_{1}(t)$ represent
the real phases of $c_{0}(t)$ and $c_{1}(t)$, respectively. Next, employing
the complex quantum amplitudes $c_{0}(t)$ and $c_{1}(t)$ in Eq. (\ref{qa}),
rewrite the state $\left\vert \psi(t)\right\rangle $ by means of a physically
equivalent state recast in its standard Bloch sphere representation
characterized by the polar angle $\theta\left(  t\right)  \in\left[  0\text{,
}\pi\right]  $ and the azimuthal angle $\varphi\left(  t\right)  \in\left[
0\text{, }2\pi\right)  $. From the knowledge of the temporal change of the two
spherical angles $\theta\left(  t\right)  $ and $\varphi\left(  t\right)  $,
define the volume of the parametric region accessed by the quantum-mechanical
system during its evolution from $\left\vert \psi(t_{A})\right\rangle
=\left\vert A\right\rangle $ to $\left\vert \psi(t)\right\rangle $. Lastly,
evaluate the temporal-average volume of the parametric region accessed by the
quantum-mechanical system during its evolution from $\left\vert \psi
(t_{A})\right\rangle =\left\vert A\right\rangle $ to $\left\vert \psi
(t_{B})\right\rangle =\left\vert B\right\rangle $ with $t\in\left[
t_{A}\text{, }t_{B}\right]  $.

Following this initial outline, we will now delve into the specifics of the
calculation process for $\overline{\mathrm{V}}\left(  t_{A}\text{, }%
t_{B}\right)  $. Employing Eq. (\ref{qa}), we note that $\left\vert
\psi(t)\right\rangle =c_{0}(t)\left\vert 0\right\rangle +c_{1}(t)\left\vert
1\right\rangle $ is physically equivalent to the state $\left\vert
c_{0}(t)\right\vert \left\vert 0\right\rangle +\left\vert c_{1}(t)\right\vert
e^{i\left[  \phi_{1}(t)-\phi_{0}(t)\right]  }\left\vert 1\right\rangle $.
Therefore, $\left\vert \psi(t)\right\rangle $ can be rewritten
\begin{equation}
\left\vert \psi(t)\right\rangle =\cos\left(  \frac{\theta\left(  t\right)
}{2}\right)  \left\vert 0\right\rangle +e^{i\varphi\left(  t\right)  }%
\sin\left(  \frac{\theta\left(  t\right)  }{2}\right)  \left\vert
1\right\rangle \text{.} \label{qa2}%
\end{equation}
From a formal standpoint, the polar angle $\theta\left(  t\right)  $ and the
azimuthal angle $\varphi\left(  t\right)  \overset{\text{def}}{=}\phi
_{1}(t)-\phi_{0}(t)=\arg\left[  c_{1}(t)\right]  -\arg\left[  c_{0}(t)\right]
$ in Eq. (\ref{qa2}) can be defined as%
\begin{equation}
\theta\left(  t\right)  \overset{\text{def}}{=}2\arctan\left(  \frac
{\left\vert c_{1}(t)\right\vert }{\left\vert c_{0}(t)\right\vert }\right)
\text{,} \label{teta}%
\end{equation}
and, taking as working conditions $\operatorname{Re}\left[  c_{1}(t)\right]
>0$ and $\operatorname{Re}\left[  c_{0}(t)\right]  >0$,
\begin{equation}
\varphi\left(  t\right)  \overset{\text{def}}{=}\arctan\left\{  \frac
{\operatorname{Im}\left[  c_{1}(t)\right]  }{\operatorname{Re}\left[
c_{1}(t)\right]  }\right\}  -\arctan\left\{  \frac{\operatorname{Im}\left[
c_{0}(t)\right]  }{\operatorname{Re}\left[  c_{0}(t)\right]  }\right\}
\text{,} \label{fi}%
\end{equation}
respectively. In general, however, the functional form for $\varphi\left(
t\right)  $ in Eq. (\ref{fi}) can assume a more convoluted expression. This is
caused by the fact that, in general, the phase $\arg\left(  z\right)  $ of a
complex number $%
\mathbb{C}
\ni z\overset{\text{def}}{=}x+iy=\left\vert z\right\vert e^{i\arg(z)}$ should
be expressed in terms of the $2$-\textrm{argument arctangent} function
\textrm{atan}$2$ as $\arg(z)=$ \textrm{atan}$2(y$, $x)$. If $x>0$,
\textrm{atan}$2(y$, $x)$ becomes $\arctan\left(  y/x\right)  $. For additional
mathematical details on \textrm{atan}$2$, we suggest Ref. \cite{grad00}. So,
given $\theta\left(  t\right)  $ and $\varphi\left(  t\right)  $, one notes
that the unit three-dimensional Bloch vector $\hat{r}\left(  t\right)  $ that
corresponds to the state vector $\left\vert \psi(t)\right\rangle $ in Eq.
(\ref{qa2}) is equal to $\hat{r}\left(  t\right)  =(\sin\left[  \theta\left(
t\right)  \right]  \cos\left[  \varphi\left(  t\right)  \right]  $,
$\sin\left[  \theta\left(  t\right)  \right]  \sin\left[  \varphi\left(
t\right)  \right]  $, $\cos\left[  \theta\left(  t\right)  \right]  )$. We are
currently able to establish a definition for $\overline{\mathrm{V}}\left(
t_{A}\text{, }t_{B}\right)  $. Specifically, the accessed volume
$\overline{\mathrm{V}}\left(  t_{A}\text{, }t_{B}\right)  $ that corresponds
to the quantum evolution governed by the Hamiltonian \textrm{H}$\left(
t\right)  $ from $\left\vert \psi(t_{A})\right\rangle =\left\vert
A\right\rangle $ to $\left\vert \psi(t_{B})\right\rangle =\left\vert
B\right\rangle $, with $t\in\left[  t_{A}\text{, }t_{B}\right]  $, is given by%
\begin{equation}
\overline{\mathrm{V}}\left(  t_{A}\text{, }t_{B}\right)  \overset{\text{def}%
}{=}\frac{1}{t_{B}-t_{A}}\int_{t_{A}}^{t_{B}}V(t)dt\text{.}
\label{avgcomplexity}%
\end{equation}
The quantity $V(t)$ that appears in Eq. (\ref{avgcomplexity}) denotes the
instantaneous volume defined as,%
\begin{equation}
V(t)=V(\theta(t)\text{, }\varphi\left(  t\right)  )\overset{\text{def}}%
{=}\mathrm{vol}\left[  \mathcal{D}_{\mathrm{accessed}}\left[  \theta(t)\text{,
}\varphi(t)\right]  \right]  \text{,} \label{local-complexity}%
\end{equation}
where $\mathrm{vol}\left[  \mathcal{D}_{\mathrm{accessed}}\left[
\theta(t)\text{, }\varphi(t)\right]  \right]  $ is given by,%
\begin{equation}
\mathrm{vol}\left[  \mathcal{D}_{\mathrm{accessed}}\left[  \theta(t)\text{,
}\varphi(t)\right]  \right]  \overset{\text{def}}{=}\int\int_{\mathcal{D}%
_{\mathrm{accessed}}\left[  \theta(t)\text{, }\varphi(t)\right]  }%
\sqrt{g_{\mathrm{FS}}\left(  \theta\text{, }\varphi\right)  }d\theta
d\varphi\text{.} \label{q3}%
\end{equation}
We remark that since we consider these volumes to be specified by positive
real numerical values, $\mathrm{vol}\left[  \mathcal{\cdot}\right]  $ means
$\left\vert \mathrm{vol}\left[  \mathcal{\cdot}\right]  \right\vert \geq0$. In
Eq. (\ref{q3}), $g_{\mathrm{FS}}\left(  \theta\text{, }\varphi\right)
\overset{\text{def}}{=}\sqrt{\sin^{2}(\theta)/16}$ denotes the determinant of
the matrix associated with the Fubini-Study infinitesimal line element
$ds_{\mathrm{FS}}^{2}\overset{\text{def}}{=}(1/4)\left[  d\theta^{2}+\sin
^{2}(\theta)d\varphi^{2}\right]  $. Lastly, $\mathcal{D}_{\mathrm{accessed}%
}\left[  \theta(t)\text{, }\varphi(t)\right]  $ in Eq. (\ref{q3}) defines the
parametric region accessed by the quantum-mechanical system during its
evolution from the initial state $\left\vert \psi(t_{A})\right\rangle
=\left\vert A\right\rangle $ to an intermediate state $\left\vert
\psi(t)\right\rangle $, with $t\in\left[  t_{A}\text{, }t_{B}\right]  $. It is
characterized as%
\begin{equation}
\mathcal{D}_{\mathrm{accessed}}\left[  \theta(t)\text{, }\varphi(t)\right]
\overset{\text{def}}{=}\left[  \theta\left(  t_{A}\right)  \text{, }%
\theta\left(  t\right)  \right]  \times\left[  \varphi\left(  t_{A}\right)
\text{, }\varphi\left(  t\right)  \right]  \subset\left[  0\text{, }%
\pi\right]  _{\theta}\times\left[  0\text{, }2\pi\right)  _{\varphi}\text{.}
\label{j5B}%
\end{equation}
For the sake of computational efficiency, we note that the instantaneous
volume $V(t)$ in Eq. (\ref{local-complexity}) can be straightforwardly recast
as $V(t)=\left\vert \left(  \cos\left[  \theta\left(  t_{A}\right)  \right]
-\cos\left[  \theta\left(  t\right)  \right]  \right)  \left(  \varphi
(t)-\varphi(t_{A})\right)  \right\vert /4$ where $\theta\left(  t\right)  $
and $\varphi\left(  t\right)  $ appear in Eqs. (\ref{teta}) and (\ref{fi}),
respectively. In summary, drawing from our traditional probabilistic framework
(in which the tilde symbol signifies the time-average process), the accessed
volume $\overline{\mathrm{V}}\left(  t_{A}\text{, }t_{B}\right)  $ in Eq.
(\ref{avgcomplexity}) can be rewritten as%
\begin{equation}
\overline{\mathrm{V}}\left(  t_{A}\text{, }t_{B}\right)  \overset{\text{def}%
}{=}\widetilde{\mathrm{vol}}\left[  \mathcal{D}_{\mathrm{accessed}}\left[
\theta(t)\text{, }\varphi(t)\right]  \right]  \text{,} \label{cafe1}%
\end{equation}
where $t\in\left[  t_{A}\text{, }t_{B}\right]  $ in Eq. (\ref{cafe1}).

To define the \emph{accessible volume} $\mathrm{V}_{\max}\left(  t_{A}\text{,
}t_{B}\right)  $, it is essential to provide some preliminary observations.
The shift from the traditional probabilistic framework, particularly regarding
the asymptotic temporal behavior of the complexity measure and the constraints
on the globally accessible parametric regions, requires careful consideration.
These two elements, along with our focus on comparing the complexity of
various quantum evolutions that connect the same initial and final states,
necessitate the introduction of a quantum complexity length scale. It is
important to recognize that complexity encompasses more than mere length.
Indeed, the idea that complexity transcends mere length is recognized across
various scientific disciplines, including computer science
\cite{beizer84,gill91,ande94} and evolutionary biology \cite{bonner04}. In the
realm of computer science, for instance, when articulating the concept of
software complexity, it can be posited that complexity ought to be regarded as
a characteristic that quantifies aspects independent of a software program's
length \cite{ande94}. It is conceivable that while complexity may rise with
the magnitude of a task, this increase is typically less pronounced than the
growth in size \cite{bonner04}. Similarly, in the context of evolutionary
biology, the concept of evolutionary complexity reveals significant
discrepancies in the established size-complexity relationship. The interplay
between size (or length) and complexity can be intricate, particularly when
examining the evolution of biological entities such as small multicellular
organisms. In these cases, alterations in complexity do not inherently
necessitate corresponding changes in size. Lastly, for a thought-provoking
exploration of how complexity should correlate with size within an information
theory framework applied to physical systems, we recommend consulting Ref.
\cite{ay08}. Therefore, we propose to establish a quantum complexity length
scale based on two primary components. The first component is the actual
length of the trajectory on the Bloch sphere that links the specified initial
and final states. It is evident that different Hamiltonians produce distinct
paths with varying lengths over a finite time interval $t_{B}-t_{A}$. The
second component accounts for the fact that different Hamiltonians result in
different bounded accessible volumes of parametric regions, each of which can
be encapsulated within a corresponding larger (yet still bounded) local
accessible volume. The ratio of these two volumes can be interpreted as a
normalized (local) volume ratio, which will define the second component of our
proposed quantum complexity length scale, as elaborated in the subsequent
subsection. With these considerations in mind, we define the accessible volume
$\mathrm{V}_{\max}(t_{A}$, $t_{B})$ as%
\begin{equation}
\mathrm{V}_{\max}(t_{A}\text{, }t_{B})\overset{\text{def}}{=}\mathrm{vol}%
\left[  \mathcal{D}_{\text{\textrm{accessible}}}\left(  \theta\text{, }%
\varphi\right)  \right]  =\int\int_{\mathcal{D}_{\text{\textrm{accessible}}%
}\left(  \theta\text{, }\varphi\right)  }\sqrt{g_{\mathrm{FS}}\left(
\theta\text{, }\varphi\right)  }d\theta d\varphi\text{.} \label{j4}%
\end{equation}
The quantity $\mathcal{D}_{\text{\textrm{accessible}}}\left(  \theta\text{,
}\varphi\right)  $ that appears in Eq. (\ref{j4}) represents the (local)
maximally accessible two-dimensional parametric region during the
quantum-mechanical transition from $\left\vert \psi_{A}\left(  \theta
_{A}\text{, }\varphi_{A}\right)  \right\rangle $ to $\left\vert \psi\left(
\theta_{B}\text{, }\varphi_{B}\right)  \right\rangle $ and is given by%
\begin{equation}
\mathcal{D}_{\text{\textrm{accessible}}}\left(  \theta\text{, }\varphi\right)
\overset{\text{def}}{=}\left\{  \left(  \theta\text{, }\varphi\right)
:\theta_{\min}\leq\theta\leq\theta_{\max}\text{, and }\varphi_{\min}%
\leq\varphi\leq\varphi_{\max}\right\}  \text{.} \label{j5}%
\end{equation}
Observe that $\theta_{\min}$, $\theta_{\max}$, $\varphi_{\min}$, and
$\varphi_{\max}$ in Eq. (\ref{j5}) are defined as
\begin{equation}
\theta_{\min}\overset{\text{def}}{=}\underset{t_{A}\leq t\leq t_{B}}{\min
}\theta(t)\text{, }\theta_{\max}\overset{\text{def}}{=}\underset{t_{A}\leq
t\leq t_{B}}{\max}\theta(t)\text{, }\varphi_{\min}\overset{\text{def}}%
{=}\underset{t_{A}\leq t\leq t_{B}}{\min}\varphi(t)\text{, and }\varphi_{\max
}\overset{\text{def}}{=}\underset{t_{A}\leq t\leq t_{B}}{\max}\varphi
(t)\text{,} \label{minmax}%
\end{equation}
respectively. Moreover, note that $\mathcal{D}_{\text{\textrm{accessed}}%
}\left(  \theta\text{, }\varphi\right)  \subset\mathcal{D}%
_{\text{\textrm{accessible}}}\left(  \theta\text{, }\varphi\right)
\subset\left[  0,\pi\right]  _{\theta}\times\left[  0\text{, }2\pi\right)
_{\varphi}$. Finally, having at our disposal $\overline{\mathrm{V}}\left(
t_{A}\text{, }t_{B}\right)  $ and $\mathrm{V}_{\max}\left(  t_{A}\text{,
}t_{B}\right)  $ in Eqs. (\ref{avgcomplexity}) and (\ref{j4}), respectively,
our suggested notion of complexity \textrm{C}$\left(  t_{A}\text{, }%
t_{B}\right)  $ in Eq. (\ref{QCD}) can be defined in a formal manner. In Fig.
$1$, we propose a schematic visualization of $\mathrm{V}_{\mathrm{tot}%
}\overset{\text{def}}{=}$\textrm{vol}$\left[  \left[  0,\pi\right]  _{\theta
}\times\left[  0\text{, }2\pi\right)  _{\varphi}\right]  =\pi$, $\mathrm{V}%
_{\max}$, and $\overline{\mathrm{V}}$. \begin{figure}[t]
\centering
\includegraphics[width=0.75\textwidth] {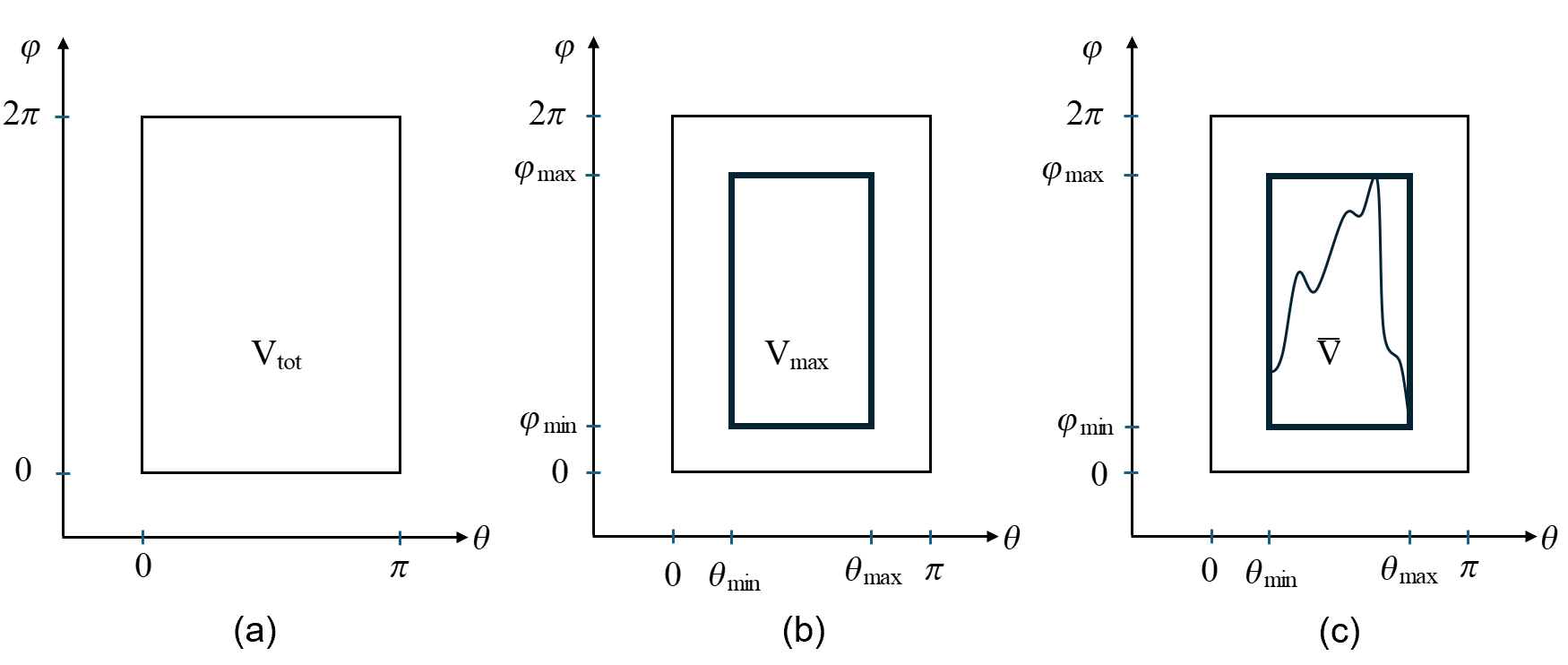}\caption{Schematic visualization
of the total volume \textrm{V}$_{\mathrm{tot}}$ in (a), the accessible volume
\textrm{V}$_{\max}$ in (b), and the accessed volume $\overline{\mathrm{V}}$ in
(c). The total volume \textrm{V}$_{\mathrm{tot}}=\mathrm{vol}(\mathcal{D}%
_{\mathrm{tot}})$ in (a) is the volume of the whole parameter space
$\mathcal{D}_{\mathrm{tot}}$ (i.e., the region bounded by the thin solid
rectangular line). The accessible volume \textrm{V}$_{\max}=\mathrm{vol}%
(\mathcal{D}_{\mathrm{accessible}})$ in (b) is the volume of the accessible
parameter space $\mathcal{D}_{\mathrm{accessible}}$ (i.e., the region bounded
by the thick solid rectangular line). Furthermore, the accessed volume
$\overline{\mathrm{V}}=\mathrm{vol}(\mathcal{D}_{\mathrm{accessed}})$ in (c)
is the volume of the accessed parameter space $\mathcal{D}_{\mathrm{accessed}%
}$ (i.e., the region bounded by the curvy solid line). Finally, note that
$\overline{\mathrm{V}}\leq$\textrm{V}$_{\max}\leq$\textrm{V}$_{\mathrm{tot}}$
since $\mathcal{D}_{\mathrm{accessed}}\subseteq\mathcal{D}%
_{\mathrm{accessible}}\subseteq\mathcal{D}_{\mathrm{tot}}$ and, in addition,
\textrm{V}$_{\max}-\overline{\mathrm{V}}=\mathrm{vol}(\mathcal{D}%
_{\mathrm{accessible}}\backslash\mathcal{D}_{\mathrm{accessed}})$ is the
volume of the fraction of accessible region not being effectively accessed
during the quantum evolution.}%
\end{figure}

It is important to highlight that within the framework of Bayesian model
selection, the term Bayesian complexity is rigorously defined as the natural
logarithm of the ratio of two volumes, expressed as \textrm{C}%
$_{\mathrm{Bayesian}}\overset{\text{def}}{=}\ln\left[  V(f)/V_{\mathrm{c}%
}(f)\right]  $ \cite{myung00}. Here, $V_{\mathrm{c}}(f)$ represents the volume
of the distinguishable distributions in $f$ that approximate the true
distribution, while $V(f)$ denotes the total volume of the model family $f$.
Notably, a small ratio of $V_{\mathrm{c}}(f)/V(f)$ indicates complex models
that occupy a limited volume near the truth in relation to the overall model
volume. Consequently, a complex model, from a Bayesian perspective, is
characterized by a minimal proportion of its distinguishable probability
distributions that are situated close to the truth \cite{myung00}. In a
similar vein, by equating $V_{\mathrm{c}}$ with $\overline{\mathrm{V}}$ and
$V$ with $\mathrm{V}_{\max}$, the ratio $\overline{\mathrm{V}}/\mathrm{V}%
_{\max}$ is introduced to facilitate a meaningful comparison of various
Hamiltonian evolutions, which may be associated with parametric regions of
differing dimensions. Additionally, the ratio $\overline{\mathrm{V}%
}/\mathrm{V}_{\max}$ serves to penalize Hamiltonian evolutions that correspond
to an accessed parametric region with a volume that constitutes only a small
fraction of the total accessible volume. Ultimately, a complex quantum
Hamiltonian evolution is defined as one with a low fraction $\overline
{\mathrm{V}}/\mathrm{V}_{\max}$, indicating a significant proportion $\left(
\mathrm{V}_{\max}-\overline{\mathrm{V}}\right)  /\mathrm{V}_{\max}$. With the
establishment of a complexity notion for quantum evolution, we can now proceed
to define the concept of complexity length scale.

We start by remembering the notion of length of a path on the Bloch sphere. To
specify the physical meaning of the Riemannian distance between two pure
quantum states chosen in an arbitrary fashion, we refer to the investigation
by Wootters in Ref. \cite{wootters81}. Alternatively, for mixed quantum
states, we refer to the work performed by Braunstein and Caves in Ref.
\cite{braunstein94}. Two infinitesimally close points $\xi$ and $\xi+d\xi$
along a path $\xi\left(  \chi\right)  $ with $\chi_{1}\leq\chi\leq\chi_{2}$
can be regarded as distinguishable from a statistical standpoint when $d\xi$
is greater than (or equal to) the standard fluctuation of $\xi$ \cite{diosi84}%
. Typically, the infinitesimal line element along the path is equal to
$ds_{\mathrm{FS}}$ with $ds_{\mathrm{FS}}^{2}=g_{\mu\nu}^{\mathrm{FS}}\left(
\xi\right)  d\xi^{\mu}d\xi^{\nu}$ with $1\leq\mu$, $\nu\leq m$, with $m$
denoting the number of real parameters employed in the parametrization of the
quantum state. Then, the length $\mathcal{L}$ of the path $\xi\left(
\chi\right)  $ between $\xi_{1}\overset{\text{def}}{=}\xi\left(  \chi
_{1}\right)  $ and $\xi_{2}\overset{\text{def}}{=}\xi\left(  \chi_{2}\right)
$ with $\chi_{1}\leq\chi\leq\chi_{2}$ is defined as%
\begin{equation}
\mathcal{L}\overset{\text{def}}{=}\int_{\xi_{1}}^{\xi_{2}}\sqrt
{ds_{\mathrm{FS}}^{2}}=\int_{\chi_{1}}^{\chi_{2}}\sqrt{\frac{ds_{\mathrm{FS}%
}^{2}}{d\chi^{2}}}d\chi\text{,}%
\end{equation}
and represents the maximal number $\tilde{N}$ of quantum states along the path
$\xi\left(  \chi\right)  $ that are distinguishable in statistical terms. In
the case of two-level quantum systems, we have $ds_{\mathrm{FS}}^{2}%
\overset{\text{def}}{=}(1/4)\left[  d\theta^{2}+\sin^{2}(\theta)d\varphi
^{2}\right]  $. The geodesic distance between $\xi_{1}$ and $\xi_{2}$ is
notably the shortest path connecting these two points, determined by the
minimum of $\tilde{N}$. This relationship between path length and the number
of statistically distinguishable states along the trajectory is an important
physical aspect to consider throughout our analysis. Furthermore, this
viewpoint can be naturally extended to the geometric characterization of
quantum mixed states \cite{braunstein94}. More specifically, from the
perspective of geodesic efficiency, it is understood that an efficient quantum
evolution corresponds to a path length $s$ as described by \cite{anandan90},%
\begin{equation}
s\overset{\text{def}}{=}\int_{t_{A}}^{t_{B}}2\frac{\Delta E\left(  t\right)
}{\hslash}dt\text{,} \label{j3}%
\end{equation}
linking an initial (i.e., $\left\vert \psi_{A}\left(  \theta_{A}\text{,
}\varphi_{A}\right)  \right\rangle $) and a final (i.e., $\left\vert \psi
_{B}\left(  \theta_{B}\text{, }\varphi_{B}\right)  \right\rangle $) quantum
pure state as small as possible (i.e., with the length as close as possible to
the shortest geodesic path length $s_{0}\overset{\text{def}}{=}2\arccos\left[
\left\vert \left\langle \psi_{A}\left(  \theta_{A}\text{, }\varphi_{A}\right)
\left\vert \psi_{B}\left(  \theta_{B}\text{, }\varphi_{B}\right)  \right.
\right\rangle \right\vert \right]  $ linking the two states). The quantity
$\Delta E\left(  t\right)  $ that appears in Eq. (\ref{j3}) represents the
energy uncertainty (i.e., the square-root of the variance of the Hamiltonian
evaluated with respect to the state $\left\vert \psi\left(  t\right)
\right\rangle $). Consequently, arranging (the performance of) quantum
evolutions based on geodesic efficiency $\eta_{\mathrm{GE}}\overset
{\text{def}}{=}s_{0}/s$ \cite{anandan90,cafaro20}) is a straightforward
process. However, one may question whether it is reasonable to assume that the
most efficient quantum evolution is achieved by traversing the (smallest)
average volume of the accessed parametric region in the least amount of time.
Contrary to what might be inferred from a classical probabilistic perspective,
it becomes evident that assessing the performance of quantum evolutions is
more subtle than initially anticipated when considering both the average
volume of the accessed parametric region (i.e., the accessed volume
$\overline{\mathrm{V}}(t_{A}$, $t_{B})$ in Eq. (\ref{avgcomplexity})) and the
maximum volume of the accessible parametric region defined by the accessible
volume \textrm{V}$_{\max}(t_{A}$, $t_{B})$ in Eq. (\ref{j4}).

In recognition that complexity encompasses more than mere length, we
quantitatively articulate this nuance by presenting a \emph{complexity length
scale} $\mathrm{L}_{\mathrm{C}}(t_{A}$, $t_{B})\geq s(t_{A}$, $t_{B})$ for any
quantum-mechanical evolution characterized by a length $s$ of the path that
connects initial and final quantum states $\left\vert \psi_{A}\left(
\theta_{A}\text{, }\varphi_{A}\right)  \right\rangle $ and $\left\vert
\psi_{B}\left(  \theta_{B}\text{, }\varphi_{B}\right)  \right\rangle $,
respectively. We propose to define $\mathrm{L}_{\mathrm{C}}$ as%
\begin{equation}
\mathrm{L}_{\mathrm{C}}(t_{A}\text{, }t_{B})\overset{\text{def}}{=}%
\frac{s(t_{A}\text{, }t_{B})}{\sqrt{\frac{\overline{\mathrm{V}}(t_{A}\text{,
}t_{B})}{\mathrm{V}_{\max}(t_{A}\text{, }t_{B})}}}\text{,} \label{complexityL}%
\end{equation}
that is, employing Eqs. (\ref{avgcomplexity}), (\ref{j3}), and (\ref{j4}),%
\begin{equation}
\mathrm{L}_{\mathrm{C}}\left(  t_{A}\text{, }t_{B}\right)  \overset
{\text{def}}{=}\left(  \int_{t_{A}}^{t_{B}}2\frac{\Delta E\left(  t\right)
}{\hslash}dt\right)  \left(  \frac{\frac{1}{t_{B}-t_{A}}\int_{t_{A}}^{t_{B}%
}\left\vert \int\int_{\mathcal{D}_{\text{\textrm{accessed}}}\left(
\theta\text{, }\varphi\right)  }\sqrt{g_{\mathrm{FS}}\left(  \theta\text{,
}\varphi\right)  }d\theta d\varphi\right\vert dt}{\left\vert \int
\int_{\mathcal{D}_{\text{\textrm{accessible}}}\left(  \theta\text{, }%
\varphi\right)  }\sqrt{g_{\mathrm{FS}}\left(  \theta\text{, }\varphi\right)
}d\theta d\varphi\right\vert }\right)  ^{-1/2}\text{.} \label{QCLS}%
\end{equation}
The quantity $\mathrm{L}_{\mathrm{C}}\left(  t_{A}\text{, }t_{B}\right)  $
presented in Eq. (\ref{QCLS}) represents the proposed quantum complexity
length scale for a quantum evolution occurring within a finite time interval
$t_{B}-t_{A}$. From the perspective of dimensional analysis, Eq.
(\ref{complexityL}) can be interpreted as arising from the informed
assumption, $s^{2}:\overline{\mathrm{V}}=\mathrm{L}_{\mathrm{C}}%
^{2}:\mathrm{V}_{\max}$. While the term \textquotedblleft
volume\textquotedblright\ has been utilized throughout this paper, it is
important to note that the parametric regions described in Eqs. (\ref{j5B})
and (\ref{j5}) are, in fact, two-dimensional. Broadly speaking, the intuitive
rationale behind our definition of the complexity quantum length scale in Eq.
(\ref{complexityL}) can be encapsulated as follows: For a specific pair of
initial and final quantum states on the Bloch sphere, a shorter connecting
path that utilizes a significant portion of the locally accessible parameter
space is considered less complex than a longer path that only engages a minor
fraction of that space. The length scale $\mathrm{L}_{\mathrm{C}}$ aims to
convey the following intuitive notion: A longer path that occupies a smaller
portion of the maximum volume of the accessible parametric region is deemed
more complex than a shorter path that occupies a larger portion of that
region. In principle, a more efficient quantum evolution does not necessarily
equate to a less complex one (i.e., $s_{1}\leq s_{2}$ does not inherently
imply $\mathrm{L}_{\mathrm{C}_{1}}\leq\mathrm{L}_{\mathrm{C}_{2}}$). In fact,
a less complex quantum evolution may not always traverse a smaller average
accessed volume in a shorter duration. Instead, it is characterized by an
average accessed volume that is as close as possible to the maximum accessible
volume. Our proposed complexity length scale incorporates both aspects: the
length of the path and the ratio of accessed to accessible volumes, thereby
reflecting the complexity of the evolution. Indeed, employing Eqs. (\ref{QCD})
and (\ref{complexityL}), a different representation of $\mathrm{L}%
_{\mathrm{C}}$ becomes%
\begin{equation}
\mathrm{L}_{\mathrm{C}}(t_{A}\text{, }t_{B})=\frac{s(t_{A}\text{, }t_{B}%
)}{\sqrt{1-\mathrm{C}\left(  t_{A}\text{, }t_{B}\right)  }}\text{.}
\label{lisa}%
\end{equation}
Inspecting Eq. (\ref{lisa}), we observe that $\mathrm{L}_{\mathrm{C}}(t_{A}$,
$t_{B})\geq s(t_{A}$, $t_{B})$ since $0\leq\mathrm{C}\left(  t_{A}\text{,
}t_{B}\right)  \leq1$ by construction. Moreover, it is worth noting
that\textbf{ }$s_{2}(t_{A}$, $t_{B})\geq s_{1}(t_{A}$, $t_{B})$ does not lead
automatically to $\mathrm{L}_{\mathrm{C}_{2}}(t_{A}$, $t_{B})\geq
\mathrm{L}_{\mathrm{C}_{1}}(t_{A}$, $t_{B})$. As a matter of fact, when
$\mathrm{C}_{2}\left(  t_{A}\text{, }t_{B}\right)  $ is sufficiently smaller
than $\mathrm{C}_{1}\left(  t_{A}\text{, }t_{B}\right)  $ and, additionally,
$s_{2}(t_{A}$, $t_{B})$ is not too much greater than $s_{1}(t_{A}$, $t_{B})$,
we are still able to possess $\mathrm{L}_{\mathrm{C}_{2}}(t_{A}$, $t_{B}%
)\leq\mathrm{L}_{\mathrm{C}_{1}}(t_{A}$, $t_{B})$. The complexity length scale
associated with quantum evolutions that result in longer paths may be less
than that of evolutions resulting in shorter paths, given that the volumes
accessed by each are nearly equivalent to their respective accessible volumes.
In other words, it is desirable for $\mathrm{C}\left(  t_{A}\text{, }%
t_{B}\right)  $ to be minimized. In Fig. $2$, we sketch a drawing that aims to
pictorially capture the essential meaning of our proposed notions of
complexity $\mathrm{C}$ and complexity length scale $\mathrm{L}_{\mathrm{C}}%
$.\begin{figure}[t]
\centering
\includegraphics[width=0.75\textwidth] {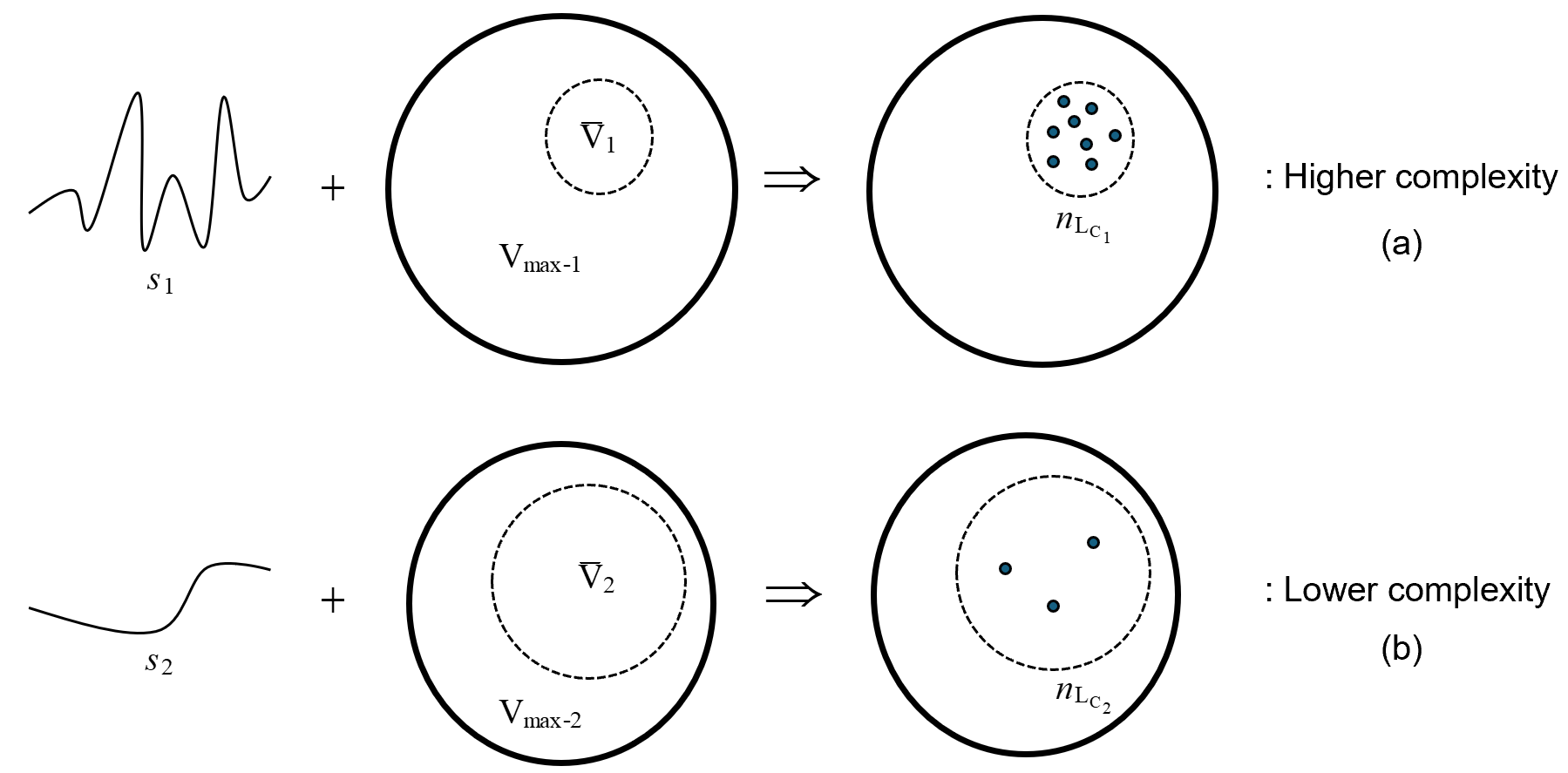}\caption{Pictorial representation
that helps explain the notion of complexity \textrm{C}. In (a), we display a
complex scenario where the accessed volume $\overline{\mathrm{V}}_{1}%
$\textrm{\ }(region bounded by the dashed circular line) is a small fraction
of the accessible volume \textrm{V}$_{\max\text{-}1}$ (region bounded by the
thick solid circular line). In (b), instead, we represent a scenario with a
smaller degree of complexity since $\overline{\mathrm{V}}_{2}$ (region bounded
by the dashed circular line) is a larger fraction of \textrm{V}$_{\max
\text{-}2}$ (region bounded by the thick solid circular line). The length $s$
of a path provides an estimate of the maximal number $n_{s}$ of statistically
distinguishable states along the path. In general, to the higher complexity
scenario in (a) there corresponds a longer quantum path of length $s_{1}$
characterized by a small fraction ($\overline{\mathrm{V}}_{1}/\mathrm{V}%
_{\max\text{-}1}$) of the accessible volume \textrm{V}$_{\max\text{-}1}$. To
the lower complexity scenario in (b), instead, there generally corresponds a
shorter quantum path of length $s_{2}<s_{1}$ described by a large fraction
($\overline{\mathrm{V}}_{2}/\mathrm{V}_{\max\text{-}2}$, with $\overline
{\mathrm{V}}_{2}/\mathrm{V}_{\max\text{-}2}>\overline{\mathrm{V}}%
_{1}/\mathrm{V}_{\max\text{-}1}$) of the accessible volume \textrm{V}%
$_{\max\text{-}2}$. The number $n_{\mathrm{L}_{\mathrm{C}}}$ of points
depicted in an accessed volume $\overline{\mathrm{V}}$ is proportional to the
squared length $s^{2}$ of the path boosted by a factor which is the reciprocal
of $\overline{\mathrm{V}}/\mathrm{V}_{\max}$ with $0\leq$ $\overline
{\mathrm{V}}/\mathrm{V}_{\max}\leq1$. These two scenarios in (a) and (b) are
properly characterized by the concept of complexity length scale
\textrm{L}$_{\mathrm{C}}$, with $n_{\mathrm{L}_{\mathrm{C}}}\overset
{\text{def}}{=}$\textrm{L}$_{\mathrm{C}}^{2}=s^{2}\cdot(\overline{\mathrm{V}%
}/\mathrm{V}_{\max})^{-1}$.}%
\end{figure}

With the concepts of complexity $\mathrm{C}\left(  t_{A}\text{, }t_{B}\right)
$ in Eq. (\ref{QCD}) and complexity length scale $\mathrm{L}_{\mathrm{C}_{2}%
}(t_{A}$, $t_{B})$ in Eq. (\ref{complexityL}), we can now commence our crucial
validation checks designed to verify the physical relevance of the framework
that supports our proposed complexity measure.

\section{Evolution of Two-level Quantum Systems}

In this section, to conduct critical validation assessments to verify the
physical relevance of the framework that supports our proposed complexity
measure, we include a direct comparison of the numerical values linked to the
complexities of suitably selected time-optimal and time sub-optimal evolutions
from suitably chosen source and target states. The Hamiltonian evolutions that
we consider here are specified by the one-parameter $\alpha\in\left[  0\text{,
}\pi\right]  $ family of Hamiltonian operators defined as,%
\begin{equation}
\mathrm{H}\left(  \alpha\right)  \overset{\text{def}}{=}E\hat{n}\left(
\alpha\right)  \cdot\mathbf{\boldsymbol{\sigma}}\text{, } \label{lost0}%
\end{equation}
where the unit vector $\hat{n}\left(  \alpha\right)  $ is given by
\begin{equation}
\hat{n}\left(  \alpha\right)  \overset{\text{def}}{=}\cos\left(
\alpha\right)  \frac{\hat{a}+\hat{b}}{\left\Vert \hat{a}+\hat{b}\right\Vert
}+\sin(\alpha)\frac{\hat{a}\times\hat{b}}{\left\Vert \hat{a}\times\hat
{b}\right\Vert }\text{,} \label{lost}%
\end{equation}
and $\mathbf{\boldsymbol{\sigma}}$ is the Pauli vector operator. In what
follows, time optimal evolutions are specified by $\hat{n}_{\mathrm{opt}}%
=\hat{n}\left(  \alpha\right)  $ with $\alpha=\pi/2$. Furthermore, time
sub-optimal evolution Hamiltonians are characterized by $\hat{n}%
_{\mathrm{sub}\text{-}\mathrm{opt}}=\hat{n}\left(  \alpha\right)  $ with
$\alpha\neq\pi/2$. For a graphical depiction of $\hat{n}_{\mathrm{opt}}$ and
$\hat{n}_{\mathrm{sub}\text{-}\mathrm{opt}}$, we refer to Fig. $3$. For a more
in depth geometrically driven discussion on the classes of stationary qubit
Hamiltonians in Eq. (\ref{lost0}), we suggest Refs.
\cite{rossetti24A,rossetti24B}.\begin{figure}[t]
\centering
\includegraphics[width=0.45\textwidth] {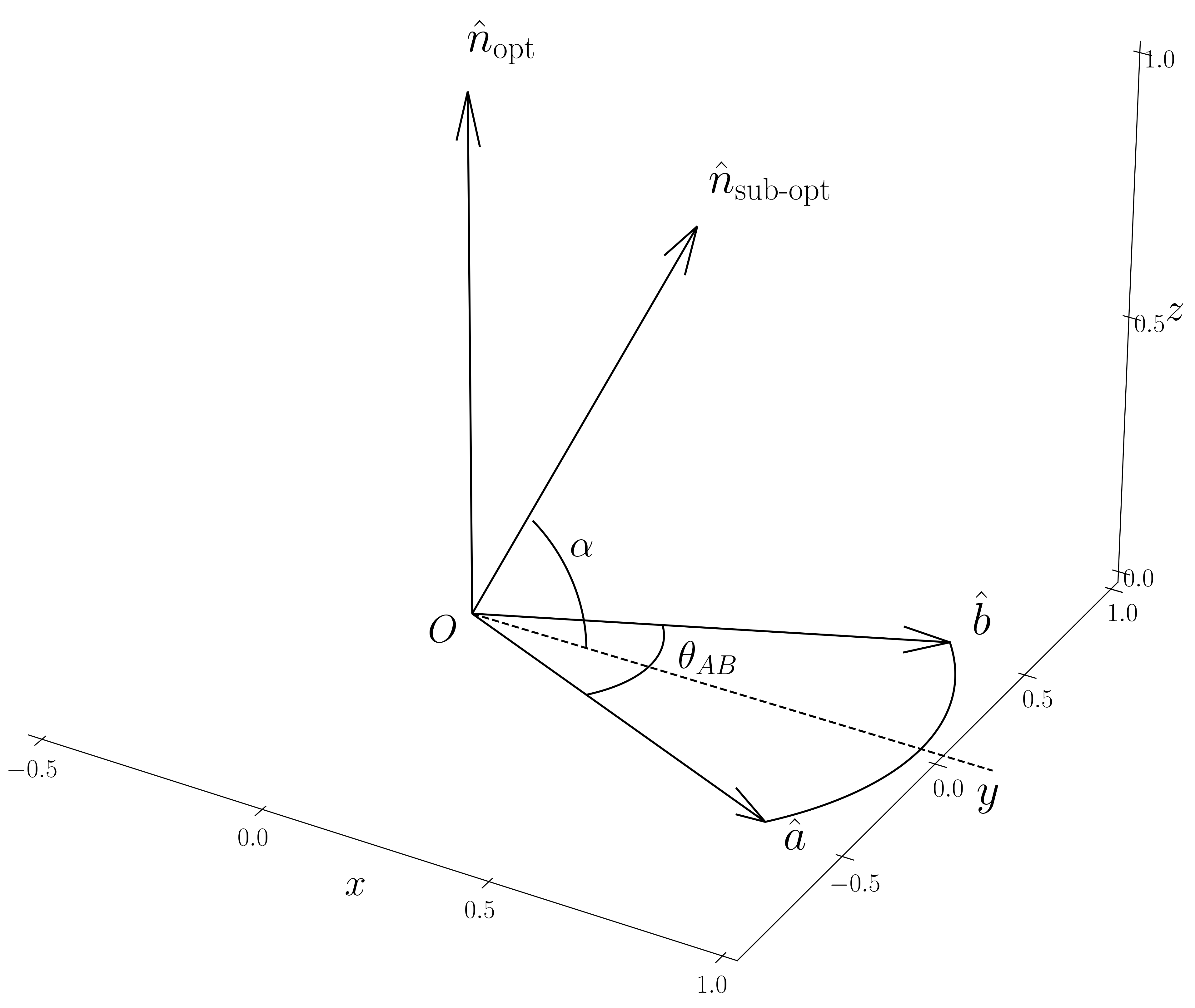}\caption{A schematic
representation of the unit vectors $\hat{n}_{\mathrm{opt}}\overset{\text{def}%
}{=}$ $(\hat{a}\times\hat{b})/\left\Vert \hat{a}\times\hat{b}\right\Vert $ and
$\hat{n}_{\mathrm{sub}\text{-}\mathrm{opt}}\overset{\text{def}}{=}\cos
(\alpha)\left[  (\hat{a}+\hat{b})/\left\Vert \hat{a}+\hat{b}\right\Vert
\right]  +\sin(\alpha)\left[  (\hat{a}\times\hat{b})/\left\Vert \hat{a}%
\times\hat{b}\right\Vert \right]  $ that illustrate the optimal and
sub-optimal rotation axes, respectively. The initial and final Bloch vectors,
denoted as $\hat{a}$ and $\hat{b}$, satisfy the condition $\hat{a}\cdot\hat
{b}=\cos(\theta_{AB})$.}%
\end{figure}

\subsection{Comparing optimal time evolutions}

We begin by investigating three distinct time-optimal evolutions not only in
terms of their complexities (i.e., the complexity \textrm{C }in Eq.
(\ref{QCD}) and the complexity length scale \textrm{L}$_{\mathrm{C}}$ in Eq.
(\ref{complexityL})), but also in terms of their geodesic efficiencies
$\eta_{\mathrm{GE}}$ in Eq. (\ref{jap}), speed efficiencies $\eta
_{\mathrm{SE}}$ in Eq. (\ref{se2}), and curvature coefficients $\kappa
_{\mathrm{AC}}^{2}$ in Eq. (\ref{XXXX}).

In this first comparative analysis, we consider time-independent Hamiltonian
evolutions specified by an Hamiltonian given by%
\begin{equation}
\mathrm{H}_{\mathrm{opt}}\overset{\text{def}}{=}E\hat{n}_{\mathrm{opt}}%
\cdot\mathbf{\boldsymbol{\sigma}}\text{, with }\hat{n}_{\mathrm{opt}}%
\overset{\text{def}}{=}\frac{\hat{a}\times\hat{b}}{\left\Vert \hat{a}%
\times\hat{b}\right\Vert }%
\end{equation}
with $\hat{a}\overset{\text{def}}{=}\hat{r}\left(  t_{i}\right)  $ and
$\hat{b}\overset{\text{def}}{=}\hat{r}\left(  t_{f}\right)  $ being the
initial and final Bloch vectors that specify the initial and final quantum
states $\left\vert \psi\left(  t_{i}\right)  \right\rangle $ and $\left\vert
\psi\left(  t_{f}\right)  \right\rangle $, respectively. Clearly, $\hat
{r}=\hat{r}\left(  t\right)  $ ($=\hat{a}\left(  t\right)  $) is the unit
Bloch vector such that $\rho\left(  t\right)  \overset{\text{def}}%
{=}\left\vert \psi\left(  t\right)  \right\rangle \left\langle \psi\left(
t\right)  \right\vert =\left[  \mathbf{1+}\hat{r}\left(  t\right)
\cdot\mathbf{\boldsymbol{\sigma}}\right]  /2$, with $\mathbf{1}$ denoting the
identity operator. Furthermore, the normalized quantum state $\left\vert
\psi\left(  t\right)  \right\rangle $ with $t_{i}\leq t\leq t_{f}$ satisfies
the time-dependent Schr\"{o}dinger evolution equation $i\hslash\partial
_{t}\left\vert \psi\left(  t\right)  \right\rangle =\mathrm{H}_{\mathrm{opt}%
}\left\vert \psi\left(  t\right)  \right\rangle $. Recall that an arbitrary
state $\left\vert \psi\left(  t\right)  \right\rangle =\left\vert \psi\left(
\theta\left(  t\right)  \text{, }\varphi\left(  t\right)  \right)
\right\rangle $ on the Bloch sphere can be recast as%
\begin{equation}
\left\vert \psi\left(  \theta\left(  t\right)  \text{, }\varphi\left(
t\right)  \right)  \right\rangle \overset{\text{def}}{=}\cos(\frac
{\theta\left(  t\right)  }{2})\left\vert 0\right\rangle +e^{i\varphi\left(
t\right)  }\sin(\frac{\theta\left(  t\right)  }{2})\left\vert 1\right\rangle
\text{,}%
\end{equation}
where $0\leq\theta\leq\pi$, $0\leq\varphi\leq2\pi$, and the corresponding unit
Bloch vector $\hat{r}\left(  t\right)  =\hat{r}\left(  \theta\left(  t\right)
\text{, }\varphi\left(  t\right)  \right)  $ is given by%
\begin{equation}
\hat{r}\left(  t\right)  \overset{\text{def}}{=}\sin\left[  \theta\left(
t\right)  \right]  \cos\left[  \varphi\left(  t\right)  \right]  \hat{x}%
+\sin\left[  \theta\left(  t\right)  \right]  \sin\left[  \varphi\left(
t\right)  \right]  \hat{y}+\cos\left[  \theta\left(  t\right)  \right]
\hat{z}\text{.}%
\end{equation}
\begin{figure}[t]
\centering
\includegraphics[width=0.35\textwidth] {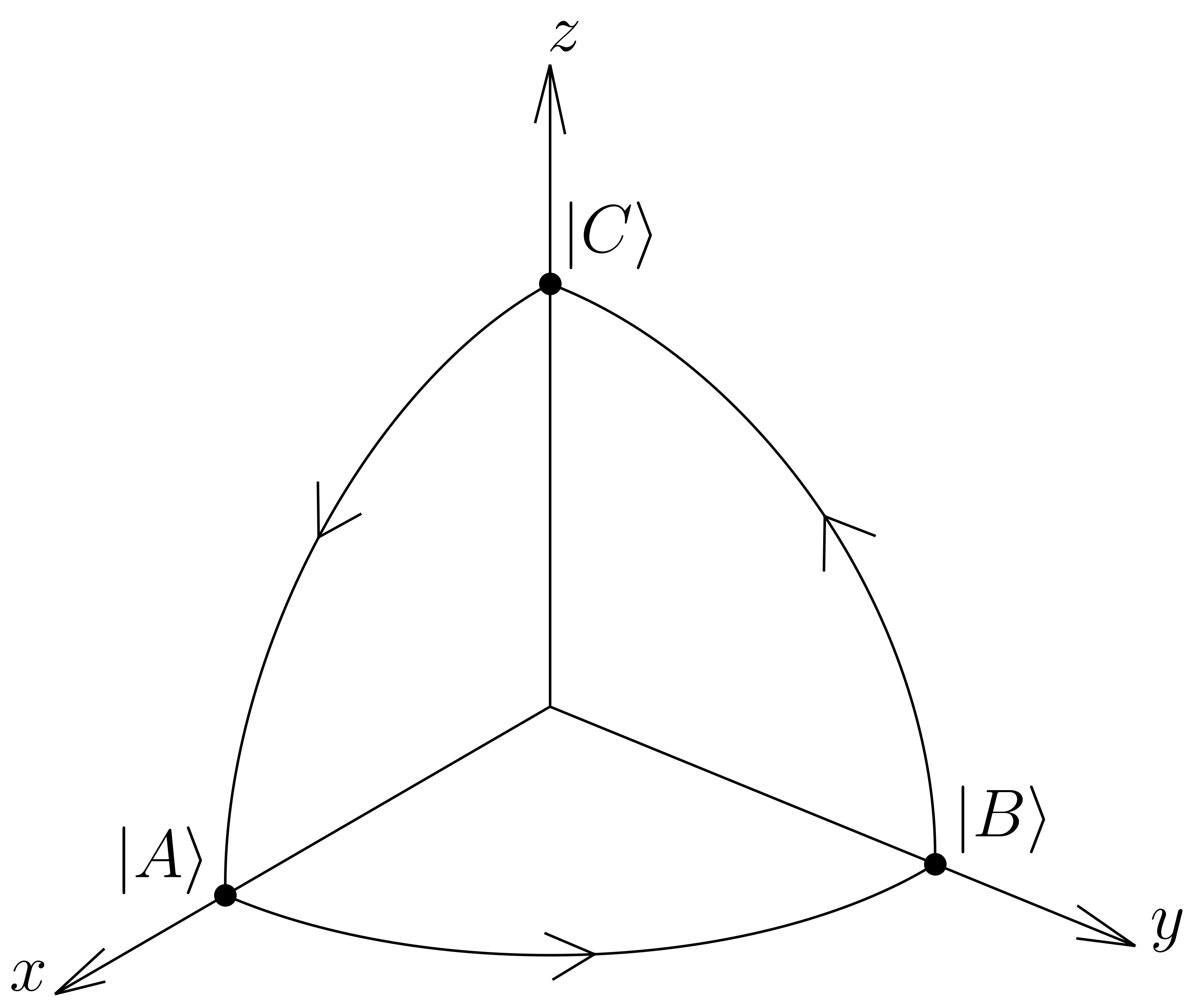}\caption{\emph{Comparing
optimal-time evolutions}. Schematic depiction of the first octant of the Bloch
sphere. The evolution from the state $\left\vert A\right\rangle \overset
{\text{def}}{=}\left[  \left\vert 0\right\rangle +\left\vert 1\right\rangle
\right]  /\sqrt{2}$ to the state $\left\vert B\right\rangle \overset
{\text{def}}{=}\left[  \left\vert 0\right\rangle +i\left\vert 1\right\rangle
\right]  /\sqrt{2}$ is governed by the Hamiltonian \textrm{H}$\overset
{\text{def}}{=}E\sigma_{z}$. Furthermore, the evolution from $\left\vert
B\right\rangle $ to $\left\vert C\right\rangle \overset{\text{def}}%
{=}\left\vert 0\right\rangle $ is specified by \textrm{H}$\overset{\text{def}%
}{=}E\sigma_{x}$. The evolution from $\left\vert C\right\rangle $ to
$\left\vert A\right\rangle $, lastly, is defined by \textrm{H}$\overset
{\text{def}}{=}E\sigma_{y}$. These three evolutions are optimal-time quantum
evolutions specified by a zero curvature coefficient and, in addition, a unit
speed efficiency. Finally, all three evolutions exhibit, as expected on
physical grounds, the same degree of complexity as specified by our proposed
measure \textrm{C}.}%
\end{figure}

\subsubsection{Connecting $\left\vert A\right\rangle $ and $\left\vert
B\right\rangle $ in Fig. $4$}

In the first evolution, we evolve the system from the state $\left\vert
A\right\rangle =\left\vert \psi\left(  t_{i}\right)  \right\rangle
\overset{\text{def}}{=}\left[  \left\vert 0\right\rangle +\left\vert
1\right\rangle \right]  /\sqrt{2}$ to the state $\left\vert B\right\rangle
=\left\vert \psi\left(  t_{f}\right)  \right\rangle \overset{\text{def}}%
{=}\left[  \left\vert 0\right\rangle +i\left\vert 1\right\rangle \right]
/\sqrt{2}$ under the Hamiltonian $\mathrm{H}_{\mathrm{opt}}\overset
{\text{def}}{=}E\hat{z}\cdot\mathbf{\boldsymbol{\sigma}}=E\sigma_{z}$. The
Bloch vectors corresponding to $\left\vert A\right\rangle $ and $\left\vert
B\right\rangle $ are given by $\hat{a}=\hat{x}\overset{\text{def}}{=}(1$, $0$,
$0)$ and $\hat{b}=\hat{y}\overset{\text{def}}{=}\left(  0\text{, }1\text{,
}0\right)  $, respectively. The state $\left\vert \psi\left(  t\right)
\right\rangle =e^{-\frac{i}{\hslash}E\sigma_{z}t}\left\vert A\right\rangle $
equals%
\begin{equation}
\left\vert \psi\left(  t\right)  \right\rangle =e^{-\frac{i}{\hslash}%
Et}\left[  \frac{1}{\sqrt{2}}\left\vert 0\right\rangle +\frac{e^{\frac
{i}{\hslash}2Et}}{\sqrt{2}}\left\vert 1\right\rangle \right]  \simeq\frac
{1}{\sqrt{2}}\left\vert 0\right\rangle +\frac{e^{\frac{i}{\hslash}2Et}}%
{\sqrt{2}}\left\vert 1\right\rangle \text{,}%
\end{equation}
with \textquotedblleft$\simeq$\textquotedblright\ denoting physical
equivalence of quantum states differing by a global phase factor. After a
simple calculation, we find that for this first evolution we have that for any
$0\leq t\leq\pi\hslash/(4E)$,
\begin{equation}
\theta(t)=\frac{\pi}{2}\text{, and }\varphi\left(  t\right)  =\frac
{2E}{\hslash}t\text{.}%
\end{equation}
It follows then that the accessed and accessible volumes, $\overline
{\mathrm{V}}$ and \textrm{V}$_{\max}$, respectively, are given by%
\begin{equation}
\overline{\mathrm{V}}=\frac{4E}{\pi\hslash}\int_{0}^{\frac{\pi\hslash}{4E}%
}\left\vert \frac{\varphi\left(  t\right)  -\varphi\left(  0\right)  }%
{2}\right\vert dt=\frac{\pi}{8}\text{, and }\mathrm{V}_{\max}=\frac{1}{2}%
\int_{0}^{\frac{\pi}{2}}d\varphi=\frac{\pi}{4}\text{,}%
\end{equation}
since $dV=(1/2)\sin\left(  \pi/2\right)  d\varphi$ and $\left(  \theta_{\min
}\text{, }\theta_{\max}\text{; }\varphi_{\min}\text{, }\varphi_{\max}\right)
=\left(  \pi/2\text{, }\pi/2\text{; }0\text{, }\pi/2\right)  $. Therefore, the
complexity of this evolution reduces to \textrm{C}$\overset{\text{def}}%
{=}(\mathrm{V}_{\max}-\overline{\mathrm{V}})/\mathrm{V}_{\max}=1/2$.
Furthermore, since $s(0$, $\pi\hslash/(4E))=2\arccos\left[  \sqrt{\left(
1+\hat{a}\cdot\hat{b}\right)  /2}\right]  =\pi/2$, we get from Eq.
(\ref{lisa}) that \textrm{L}$_{\mathrm{C}}=\pi/\sqrt{2}$.

\subsubsection{Connecting $\left\vert B\right\rangle $ and $\left\vert
C\right\rangle $ in Fig. $4$}

In the second evolution, we evolve the system from the state $\left\vert
B\right\rangle =\left\vert \psi\left(  t_{i}\right)  \right\rangle
\overset{\text{def}}{=}\left[  \left\vert 0\right\rangle +i\left\vert
1\right\rangle \right]  /\sqrt{2}$ to the state $\left\vert C\right\rangle
=\left\vert \psi\left(  t_{f}\right)  \right\rangle \overset{\text{def}}%
{=}\left\vert 0\right\rangle $ under the Hamiltonian $\mathrm{H}%
_{\mathrm{opt}}\overset{\text{def}}{=}E\hat{x}\cdot\mathbf{\boldsymbol{\sigma
}}=E\sigma_{x}$. The Bloch vectors corresponding to $\left\vert B\right\rangle
$ and $\left\vert C\right\rangle $ are given by $\hat{b}=\hat{y}%
\overset{\text{def}}{=}(0$, $1$, $0)$ and $\hat{c}=\hat{z}\overset{\text{def}%
}{=}\left(  0\text{, }0\text{, }1\right)  $, respectively. The state
$\left\vert \psi\left(  t\right)  \right\rangle =e^{-\frac{i}{\hslash}%
E\sigma_{x}t}\left\vert B\right\rangle $ equals%
\begin{equation}
\left\vert \psi\left(  t\right)  \right\rangle =\frac{\cos(\frac{E}{\hslash
}t)+\sin(\frac{E}{\hslash}t)}{\sqrt{2}}\left\vert 0\right\rangle +i\frac
{\cos(\frac{E}{\hslash}t)-\sin(\frac{E}{\hslash}t)}{\sqrt{2}}\left\vert
1\right\rangle \text{.}%
\end{equation}
After a simple calculation, we find that for this second evolution we have
that for any $0\leq t\leq\pi\hslash/(4E)$,
\begin{equation}
\theta(t)=2\arctan\left[  \sqrt{\left(  \frac{\cos\left(  \frac{E}{\hslash
}t\right)  -\sin\left(  \frac{E}{\hslash}t\right)  }{\cos\left(  \frac
{E}{\hslash}t\right)  +\sin\left(  \frac{E}{\hslash}t\right)  }\right)  ^{2}%
}\right]  \text{, and }\varphi\left(  t\right)  =\frac{\pi}{2}\text{.}%
\end{equation}
It follows then that the accessed and accessible volumes, $\overline
{\mathrm{V}}$ and \textrm{V}$_{\max}$, respectively, are given by%
\begin{align}
\overline{\mathrm{V}}  &  =\frac{4E}{\pi\hslash}\int_{0}^{\frac{\pi\hslash
}{4E}}\left\vert \frac{\theta\left(  t\right)  -\theta\left(  0\right)  }%
{2}\right\vert dt\nonumber\\
&  =\frac{4}{\pi}\int_{0}^{\frac{\pi}{4}}\left\vert \frac{1}{2}\left[
2\arctan\left(  \sqrt{\left(  \frac{\cos\left(  t\right)  -\sin\left(
t\right)  }{\cos\left(  t\right)  +\sin\left(  t\right)  }\right)  ^{2}%
}\right)  -\frac{\pi}{2}\right]  \right\vert dt\nonumber\\
&  =\frac{\pi}{8}\text{,}%
\end{align}
and%
\begin{equation}
\mathrm{V}_{\max}=\frac{1}{2}\int_{0}^{\frac{\pi}{2}}d\theta=\frac{\pi}%
{4}\text{,}%
\end{equation}
since $dV=(1/2)d\theta$ and $\left(  \theta_{\min}\text{, }\theta_{\max
}\text{; }\varphi_{\min}\text{, }\varphi_{\max}\right)  =\left(  0\text{, }%
\pi/2\text{; }\pi/2\text{, }\pi/2\right)  $. Therefore, the complexity of this
evolution reduces to \textrm{C}$\overset{\text{def}}{=}(\mathrm{V}_{\max
}-\overline{\mathrm{V}})/\mathrm{V}_{\max}=1/2$. Moreover, since $s(0$,
$\pi\hslash/(4E))=2\arccos\left[  \sqrt{\left(  1+\hat{b}\cdot\hat{c}\right)
/2}\right]  =\pi/2$, we get from Eq. (\ref{lisa}) that \textrm{L}%
$_{\mathrm{C}}=\pi/\sqrt{2}$.

It is important to highlight that the optimal-time propagations examined in
our study pertain to transitions between the poles and the equator of the
Bloch sphere, and vice versa. Nevertheless, a question arises regarding the
implications of considering a complete circular path, wherein the notion of
geometric Berry phase becomes relevant \cite{berry84}. From the standpoint of
geodesic efficiency, it is established that the geodesic trajectories on the
Bloch sphere represent null geometric phase curves, as noted in Ref.
\cite{cafaro23}. Investigating the complexity dynamics along both optimal and
sub-optimal circular paths that link coinciding initial and final states would
be a valuable endeavor. We will defer this exploration to future research initiatives.

\subsubsection{Connecting $\left\vert C\right\rangle $ and $\left\vert
A\right\rangle $ in Fig. $4$}

In the third evolution, we evolve the system from the state $\left\vert
C\right\rangle =\left\vert \psi\left(  t_{i}\right)  \right\rangle
\overset{\text{def}}{=}\left\vert 0\right\rangle $ to the state $\left\vert
A\right\rangle =\left\vert \psi\left(  t_{f}\right)  \right\rangle
\overset{\text{def}}{=}\left[  \left\vert 0\right\rangle +\left\vert
1\right\rangle \right]  /\sqrt{2}$ under the Hamiltonian $\mathrm{H}%
_{\mathrm{opt}}\overset{\text{def}}{=}E\hat{y}\cdot\mathbf{\boldsymbol{\sigma
}}=E\sigma_{y}$. The Bloch vectors corresponding to $\left\vert C\right\rangle
$ and $\left\vert A\right\rangle $ are given by $\hat{c}=\hat{z}%
\overset{\text{def}}{=}(0$, $0$, $1)$ and $\hat{a}=\hat{x}\overset{\text{def}%
}{=}\left(  1\text{, }0\text{, }0\right)  $, respectively. The state
$\left\vert \psi\left(  t\right)  \right\rangle =e^{-\frac{i}{\hslash}%
E\sigma_{y}t}\left\vert C\right\rangle $ equals%
\begin{equation}
\left\vert \psi\left(  t\right)  \right\rangle =\cos(\frac{E}{\hslash
}t)\left\vert 0\right\rangle +\sin(\frac{E}{\hslash}t)\left\vert
1\right\rangle \text{.}%
\end{equation}
After a simple calculation, we find that for this second evolution we have
that for any $0\leq t\leq\pi\hslash/(4E)$,
\begin{equation}
\theta(t)=2\arctan\left[  \sqrt{\frac{\sin^{2}\left(  \frac{E}{\hslash
}t\right)  }{\cos^{2}\left(  \frac{E}{\hslash}t\right)  }}\right]  \text{, and
}\varphi\left(  t\right)  =0\text{.}%
\end{equation}
It follows then that the accessed and accessible volumes, $\overline
{\mathrm{V}}$ and \textrm{V}$_{\max}$, respectively, are given by%
\begin{align}
\overline{\mathrm{V}}  &  =\frac{4E}{\pi\hslash}\int_{0}^{\frac{\pi\hslash
}{4E}}\left\vert \frac{\theta\left(  t\right)  -\theta\left(  0\right)  }%
{2}\right\vert dt\nonumber\\
&  =\frac{4}{\pi}\int_{0}^{\frac{\pi}{4}}\left\vert \arctan\left(  \sqrt
{\frac{\sin^{2}\left(  t\right)  }{\cos^{2}\left(  t\right)  }}\right)
\right\vert dt\nonumber\\
&  =\frac{\pi}{8}\text{,}%
\end{align}
and%
\begin{equation}
\mathrm{V}_{\max}=\frac{1}{2}\int_{0}^{\frac{\pi}{2}}d\theta=\frac{\pi}%
{4}\text{,}%
\end{equation}
since $dV=(1/2)d\theta$ and $\left(  \theta_{\min}\text{, }\theta_{\max
}\text{; }\varphi_{\min}\text{, }\varphi_{\max}\right)  =\left(  0\text{, }%
\pi/2\text{; }0\text{, }0\right)  $. Therefore, the complexity of this
evolution reduces to \textrm{C}$\overset{\text{def}}{=}(\mathrm{V}_{\max
}-\overline{\mathrm{V}})/\mathrm{V}_{\max}=1/2$. Furthermore, given that
$s(0$, $\pi\hslash/(4E))=2\arccos\left[  \sqrt{\left(  1+\hat{c}\cdot\hat
{a}\right)  /2}\right]  =\pi/2$, we get from Eq. (\ref{lisa}) that
\textrm{L}$_{\mathrm{C}}=\pi/\sqrt{2}$.

In Fig. $4$, we display the three trajectories that emerge from the three
optimal-time quantum evolutions that we have compared in this paper. In Table
I, instead, we report the numerical values of the geodesic efficiency
$\eta_{\mathrm{GE}}$ in Eq. (\ref{jap}), the speed efficiency $\eta
_{\mathrm{SE}}$ in Eq. (\ref{se2}), the curvature coefficient $\kappa
_{\mathrm{AC}}^{2}$ in Eq. (\ref{XXXX}), the complexity \textrm{C }in Eq.
(\ref{QCD}) and, lastly, the complexity length scale \textrm{L}$_{\mathrm{C}}$
in Eq. (\ref{complexityL}) for three time optimal quantum evolutions being
considered in this first comparative analysis of optimal time evolutions. For
completeness, we recall that to evaluate efficiencies and curvature
coefficients of the various quantum evolutions, we only need $t_{i}$, $t_{f}$,
$\hat{a}\left(  t_{i}\right)  $, $\hat{a}\left(  t_{f}\right)  $, $\hat
{a}\left(  t\right)  $ with $t_{i}\leq t\leq t_{f}$, and the time-independent
\textquotedblleft magnetic\textquotedblright\ field vector $\mathbf{h}$. More
specifically, we have $\eta_{\mathrm{GE}}=\eta_{\mathrm{GE}}\left(
t_{i}\text{, }t_{f}\text{, }\hat{a}\left(  t_{i}\right)  \text{, }\hat
{a}\left(  t_{f}\right)  \text{, }\mathbf{h}\right)  $, $\eta_{\mathrm{SE}%
}=\eta_{\mathrm{SE}}\left(  \hat{a}\left(  t\right)  \text{, }\mathbf{h}%
\right)  $, and $\kappa_{\mathrm{AC}}^{2}=\kappa_{\mathrm{AC}}^{2}\left(
\hat{a}\left(  t_{i}\right)  \text{, }\mathbf{h}\right)  =\kappa_{\mathrm{AC}%
}^{2}\left(  \hat{a}\left(  t_{f}\right)  \text{, }\mathbf{h}\right)
=\kappa_{\mathrm{AC}}^{2}\left(  \hat{a}\left(  t\right)  \text{, }%
\mathbf{h}\right)  $.\begin{table}[t]
\centering
\begin{tabular}
[c]{c|c|c|c|c|c|c|c|c}\hline\hline
\textbf{Hamiltonian} & \textbf{Initial state} & \textbf{Final state} &
\textbf{Travel time} & $\eta_{\mathrm{GE}}$ & $\eta_{\mathrm{SE}}$ &
$\kappa_{\mathrm{AC}}^{2}$ & \textrm{C} & \textrm{L}$_{\mathrm{C}}$\\\hline
$E\sigma_{z}$, time optimal & $\frac{\left\vert 0\right\rangle +\left\vert
1\right\rangle }{\sqrt{2}}$ & $\frac{\left\vert 0\right\rangle +i\left\vert
1\right\rangle }{\sqrt{2}}$ & $\frac{\pi\hslash}{4E}$ & $1$ & $1$ & $0$ &
$\frac{1}{2}$ & $\frac{\pi}{\sqrt{2}}$\\\hline
$E\sigma_{x}$, time optimal & $\frac{\left\vert 0\right\rangle +i\left\vert
1\right\rangle }{\sqrt{2}}$ & $\left\vert 0\right\rangle $ & $\frac{\pi
\hslash}{4E}$ & $1$ & $1$ & $0$ & $\frac{1}{2}$ & $\frac{\pi}{\sqrt{2}}%
$\\\hline
$E\sigma_{y}$, time optimal & $\left\vert 0\right\rangle $ & $\frac{\left\vert
0\right\rangle +\left\vert 1\right\rangle }{\sqrt{2}}$ & $\frac{\pi\hslash
}{4E}$ & $1$ & $1$ & $0$ & $\frac{1}{2}$ & $\frac{\pi}{\sqrt{2}}$\\\hline
\end{tabular}
\caption{\emph{Comparing optimal-time evolutions}. Numerical values of the
geodesic efficiency $\eta_{\mathrm{GE}}$, the speed efficiency $\eta
_{\mathrm{SE}}$, the curvature coefficient $\kappa_{\mathrm{AC}}^{2}$, the
complexity \textrm{C} and, lastly, the complexity length scale \textrm{L}%
$_{\mathrm{C}}$ for three time optimal quantum evolutions. The first evolution
in the tabular summary, for instance, occurs from the initial state $\left(
\left\vert 0\right\rangle +\left\vert 1\right\rangle \right)  /\sqrt{2}$ to
the final state $\left(  \left\vert 0\right\rangle +i\left\vert 1\right\rangle
\right)  /\sqrt{2}$ under the Hamiltonian $E\sigma_{z}$ during a (minimal)
time interval given by $\pi\hslash/\left(  4E\right)  $. As expected on
physical grounds, all three evolutions perform identically in terms of the
measures $\eta_{\mathrm{GE}}$, $\eta_{\mathrm{SE}}$, $\kappa_{\mathrm{AC}}%
^{2}$, \textrm{C}, and \textrm{L}$_{\mathrm{C}}$.}%
\end{table}

Having compared our three optimal time evolutions, we are ready for our
comparative analysis of both optimal and sub-optimal time evolutions.

\subsection{Comparing optimal and sub-optimal time evolutions}

In our first comparative analysis, we verified that the complexity of three
distinct optimal time evolutions (i.e., $E\sigma_{z}$, $E\sigma_{x}$, and
$E\sigma_{y}$) connecting three distinct pairs of initial and final states
(i.e., $\left(  \left\vert A\right\rangle \text{, }\left\vert B\right\rangle
\right)  $, $\left(  \left\vert B\right\rangle \text{, }\left\vert
C\right\rangle \right)  $, and $\left(  \left\vert C\right\rangle \text{,
}\left\vert A\right\rangle \right)  $) exhibited the same degree of
complexity, provided that the Fubini-Study distance between the pairs of
initial-final states in each one of the three distinct pairs was the same
(i.e., $\mathrm{dist}_{\mathrm{FS}}\left(  \left\vert A\right\rangle \text{,
}\left\vert B\right\rangle \right)  =\mathrm{dist}_{\mathrm{FS}}\left(
\left\vert B\right\rangle \text{, }\left\vert C\right\rangle \right)
=\mathrm{dist}_{\mathrm{FS}}\left(  \left\vert C\right\rangle \text{,
}\left\vert A\right\rangle \right)  $). Recall that $\mathrm{dist}%
_{\mathrm{FS}}\left(  \left\vert A\right\rangle \text{, }\left\vert
B\right\rangle \right)  \overset{\text{def}}{=}\arccos\left[  \left\vert
\left\langle A\left\vert B\right.  \right\rangle \right\vert \right]  $.

Here, in our second comparative analysis, given two distinct pairs of
initial-final states $\left(  \left\vert A\right\rangle \text{, }\left\vert
B\right\rangle \right)  $ and $\left(  \left\vert C\right\rangle \text{,
}\left\vert D\right\rangle \right)  $ with $\mathrm{dist}_{\mathrm{FS}}\left(
\left\vert A\right\rangle \text{, }\left\vert B\right\rangle \right)
=\mathrm{dist}_{\mathrm{FS}}\left(  \left\vert C\right\rangle \text{,
}\left\vert D\right\rangle \right)  $, we seek to evolve $\left\vert
A\right\rangle $ ($\left\vert C\right\rangle $) into $\left\vert
B\right\rangle $ ($\left\vert D\right\rangle $) both optimally and
sub-optimally. In particular, the two sub-optimal evolutions are taken so that
they possess the same degree of deviation from time optimality (i.e.,
$s_{\left\vert A\right\rangle \rightarrow\left\vert B\right\rangle
}^{\mathrm{sub}\text{-\textrm{opt}}}=s_{\left\vert C\right\rangle
\rightarrow\left\vert D\right\rangle }^{\mathrm{sub}\text{-\textrm{opt}}}%
\geq2\mathrm{dist}_{\mathrm{FS}}\left(  \left\vert C\right\rangle \text{,
}\left\vert D\right\rangle \right)  $). On physical grounds, we then expect
that the pair of optimal time evolutions, as well as the pair of sub-optimal
time evolutions, exhibit the very same level of complexity. In Fig. $5$, we
display the four trajectories that we use for our comparative analysis. The
thick paths denote geodesic trajectories, while the thin paths are nongeodesic
trajectories on the Bloch sphere.\begin{figure}[t]
\centering
\includegraphics[width=0.35\textwidth] {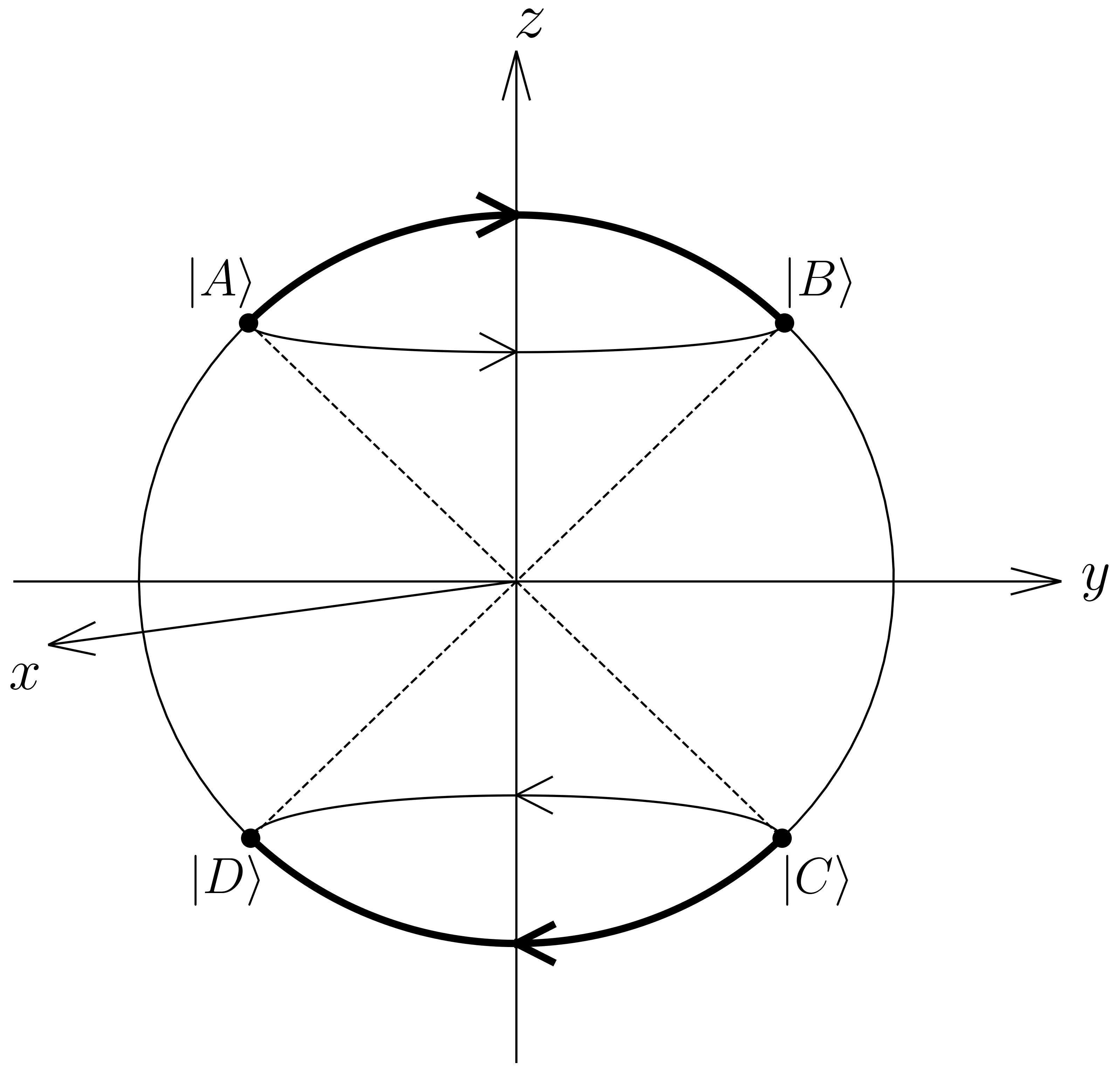}\caption{\emph{Comparing optimal
and sub-optimal time evolutions}. The (thick solid line) evolution from
$\left\vert A\right\rangle \overset{\text{def}}{=}\left[  \sqrt{2+\sqrt{2}%
}/2\right]  \left\vert 0\right\rangle -i\left[  \sqrt{2-\sqrt{2}}/2\right]
\left\vert 1\right\rangle $ to $\left\vert B\right\rangle \overset{\text{def}%
}{=}\left[  \sqrt{2+\sqrt{2}}/2\right]  \left\vert 0\right\rangle +i\left[
\sqrt{2-\sqrt{2}}/2\right]  \left\vert 1\right\rangle $ under \textrm{H}%
$\overset{\text{def}}{=}-E\sigma_{x}$ and the (thick solid line) evolution
from $\left\vert C\right\rangle \overset{\text{def}}{=}\left[  \sqrt
{2-\sqrt{2}}/2\right]  \left\vert 0\right\rangle +i\left[  \sqrt{2+\sqrt{2}%
}/2\right]  \left\vert 1\right\rangle $ to $\left\vert D\right\rangle
\overset{\text{def}}{=}\left[  \sqrt{2-\sqrt{2}}/2\right]  \left\vert
0\right\rangle -i\left[  \sqrt{2+\sqrt{2}}/2\right]  \left\vert 1\right\rangle
$ under \textrm{H}$\overset{\text{def}}{=}-E\sigma_{x}$ are optimal-time
quantum evolutions (in the $yz$-plane) specified by a zero curvature
coefficient and, in addition, a unit speed efficiency. Note that $\left\langle
A\left\vert C\right.  \right\rangle =\left\langle C\left\vert A\right.
\right\rangle =0$ and $\left\langle B\left\vert D\right.  \right\rangle
=\left\langle D\left\vert B\right.  \right\rangle =0$. Finally, both optimal
evolutions show the same degree of complexity in terms of the indicator
\textrm{C}. Alternatively, the (thin solid line) evolution from $\left\vert
A\right\rangle $ to $\left\vert B\right\rangle $ under \textrm{H}%
$\overset{\text{def}}{=}E\sigma_{z}$ and the (thin solid line) evolution from
$\left\vert C\right\rangle $ to $\left\vert D\right\rangle $ under
\textrm{H}$\overset{\text{def}}{=}-E\sigma_{z}$ are sub-optimal-time quantum
evolutions (in the $xy$-plane) specified by a nonzero curvature coefficient
and, in addition, a non-unit speed efficiency. Finally, as reasonably
expected, both sub-optimal evolutions exhibit the same level of complexity as
quantified by our proposed measure \textrm{C}.}%
\end{figure}

\subsubsection{Connecting $\left\vert A\right\rangle $ and $\left\vert
B\right\rangle $ in Fig. $5$}

In our first part of this second comparative analysis, let us assume that the
states $\left\vert A\right\rangle $ and $\left\vert B\right\rangle $ are given
by
\begin{equation}
\left\vert A\right\rangle \overset{\text{def}}{=}\frac{\sqrt{2+\sqrt{2}}}%
{2}\left\vert 0\right\rangle -i\frac{\sqrt{2-\sqrt{2}}}{2}\left\vert
1\right\rangle \text{, and }\left\vert B\right\rangle \overset{\text{def}}%
{=}\frac{\sqrt{2+\sqrt{2}}}{2}\left\vert 0\right\rangle +i\frac{\sqrt
{2-\sqrt{2}}}{2}\left\vert 1\right\rangle \text{,}%
\end{equation}
respectively. The Bloch vectors that correspond to $\left\vert A\right\rangle
$ and $\left\vert B\right\rangle $ are given by $\hat{a}\overset{\text{def}%
}{=}\left(  0\text{, }\sin\left(  -\pi/4\right)  \text{, }\cos\left(
-\pi/4\right)  \right)  =\left(  0\text{, }-1/\sqrt{2}\text{, }1/\sqrt
{2}\right)  $ and $\hat{b}\overset{\text{def}}{=}\left(  0\text{, }\sin\left(
\pi/4\right)  \text{, }\cos\left(  \pi/4\right)  \right)  =\left(  0\text{,
}1/\sqrt{2}\text{, }1/\sqrt{2}\right)  $, respectively. From these expressions
of $\hat{a}$ and $\hat{b}$, we find from Eq. (\ref{lost}) that \textrm{H}%
$\left(  \alpha\right)  $ is given by%
\begin{equation}
\mathrm{H}\left(  \alpha\right)  \overset{\text{def}}{=}E\left[  \cos\left(
\alpha\right)  \hat{z}-\sin(\alpha)\hat{x}\right]  \cdot
\mathbf{\boldsymbol{\sigma}}\text{.} \label{h1}%
\end{equation}

\paragraph{Optimal Evolution}

The optimal time evolution from $\left\vert A\right\rangle $ to $\left\vert
B\right\rangle $ occurs under the Hamiltonian \textrm{H}$_{\mathrm{opt}}%
=E\hat{n}_{\mathrm{opt}}\cdot\mathbf{\boldsymbol{\sigma}}$, with $\hat
{n}_{\mathrm{opt}}=\hat{a}\times\hat{b}/\left\Vert \hat{a}\times\hat
{b}\right\Vert =-\hat{x}$. Therefore, \textrm{H}$_{\mathrm{opt}}%
=\mathrm{H}\left(  \pi/2\right)  =-E\sigma_{x}$. After some calculations, it
turns out that the evolution from $\left\vert A\right\rangle $ to $\left\vert
B\right\rangle $ occurs from $t_{i}=0$ to $t_{f}=\pi\hslash/(4E)$ along the
state path on the Bloch sphere specified by the state $\left\vert \psi\left(
t\right)  \right\rangle =e^{\frac{i}{\hslash}E\sigma_{x}t}\left\vert
A\right\rangle $ given by%
\begin{equation}
\left\vert \psi\left(  t\right)  \right\rangle =\left[  \frac{\sqrt{2-\sqrt
{2}}}{2}\sin(\frac{E}{\hslash}t)+\frac{\sqrt{2+\sqrt{2}}}{2}\cos\left(
\frac{E}{\hslash}t\right)  \right]  \left\vert 0\right\rangle +i\left[
\frac{\sqrt{2+\sqrt{2}}}{2}\sin(\frac{E}{\hslash}t)-\frac{\sqrt{2-\sqrt{2}}%
}{2}\cos\left(  \frac{E}{\hslash}t\right)  \right]  \left\vert 1\right\rangle
\text{.}%
\end{equation}
Furthermore, after some algebra, we find that the spherical angles
$\theta\left(  t\right)  $ and $\varphi\left(  t\right)  $ with $\left\vert
\psi\left(  t\right)  \right\rangle =\left\vert \psi\left(  \theta\left(
t\right)  \text{, }\varphi\left(  t\right)  \right)  \right\rangle $ are given
by%
\begin{equation}
\theta\left(  t\right)  =2\arctan\left(  \frac{\frac{\sqrt{2+\sqrt{2}}}{2}%
\sin(\frac{E}{\hslash}t)-\frac{\sqrt{2-\sqrt{2}}}{2}\cos\left(  \frac
{E}{\hslash}t\right)  }{\frac{\sqrt{2-\sqrt{2}}}{2}\sin(\frac{E}{\hslash
}t)+\frac{\sqrt{2+\sqrt{2}}}{2}\cos\left(  \frac{E}{\hslash}t\right)
}\right)  \text{, and }\varphi\left(  t\right)  =\frac{\pi}{2}\text{,}%
\end{equation}
respectively. It follows then that the accessed and accessible volumes,
$\overline{\mathrm{V}}$ and \textrm{V}$_{\max}$, respectively, are given by%
\begin{align}
\overline{\mathrm{V}}  &  =\frac{4E}{\pi\hslash}\int_{0}^{\frac{\pi\hslash
}{4E}}\left\vert \frac{\theta\left(  t\right)  -\theta\left(  0\right)  }%
{2}\right\vert dt\nonumber\\
&  =\frac{4E}{\pi\hslash}\int_{0}^{\frac{\pi\hslash}{4E}}\left\vert
\arctan\left(  \frac{\frac{\sqrt{2+\sqrt{2}}}{2}\sin(\frac{E}{\hslash}%
t)-\frac{\sqrt{2-\sqrt{2}}}{2}\cos\left(  \frac{E}{\hslash}t\right)  }%
{\frac{\sqrt{2-\sqrt{2}}}{2}\sin(\frac{E}{\hslash}t)+\frac{\sqrt{2+\sqrt{2}}%
}{2}\cos\left(  \frac{E}{\hslash}t\right)  }\right)  +\frac{\pi}{8}\right\vert
dt\nonumber\\
&  =\frac{4}{\pi}\int_{0}^{\frac{\pi}{4}}\left\vert \arctan\left(  \frac
{\frac{\sqrt{2+\sqrt{2}}}{2}\sin(\xi)-\frac{\sqrt{2-\sqrt{2}}}{2}\cos\left(
\xi\right)  }{\frac{\sqrt{2-\sqrt{2}}}{2}\sin(\xi)+\frac{\sqrt{2+\sqrt{2}}}%
{2}\cos\left(  \xi\right)  }\right)  +\frac{\pi}{8}\right\vert d\xi\nonumber\\
&  =\frac{\pi}{8}\text{,}%
\end{align}
and,%
\begin{equation}
\mathrm{V}_{\max}=\frac{1}{2}\int_{-\frac{\pi}{4}}^{\frac{\pi}{4}}%
d\theta=\frac{\pi}{4}\text{,}%
\end{equation}
since $dV=(1/2)d\theta$ and $\left(  \theta_{\min}\text{, }\theta_{\max
}\text{; }\varphi_{\min}\text{, }\varphi_{\max}\right)  =\left(  -\pi/4\text{,
}\pi/4\text{; }\pi/2\text{, }\pi/2\right)  $. Therefore, the complexity of
this evolution reduces to \textrm{C}$\overset{\text{def}}{=}(\mathrm{V}_{\max
}-\overline{\mathrm{V}})/\mathrm{V}_{\max}=1/2$. Furthermore, given that
$s(0$, $\pi\hslash/(4E))=2\arccos\left[  \sqrt{(1+\hat{a}\cdot\hat{b}%
)/2}\right]  =\pi/2$, we get from Eq. (\ref{lisa}) that \textrm{L}%
$_{\mathrm{C}}=\pi/\sqrt{2}$.

\paragraph{Sub-optimal Evolution}

For a sub-optimal time evolution from $\left\vert A\right\rangle $ to
$\left\vert B\right\rangle $, we consider the Hamiltonian \textrm{H}%
$_{\text{\textrm{sub}-}\mathrm{opt}}=\mathrm{H}\left(  0\right)  =E\sigma_{z}%
$. After some calculations, it turns out that the evolution from $\left\vert
A\right\rangle $ to $\left\vert B\right\rangle $ occurs from $t_{i}=0$ to
$t_{f}=\pi\hslash/(2E)$ along the state path on the Bloch sphere specified by
the state $\left\vert \psi\left(  t\right)  \right\rangle =e^{-\frac
{i}{\hslash}E\sigma_{z}t}\left\vert A\right\rangle $ is given by%
\begin{equation}
\left\vert \psi\left(  t\right)  \right\rangle =e^{-\frac{i}{\hslash}%
Et}\left[  \cos\left(  \frac{\pi}{8}\right)  \left\vert 0\right\rangle
+e^{i\left(  \frac{2E}{\hslash}t-\frac{\pi}{2}\right)  }\sin\left(  \frac{\pi
}{8}\right)  \left\vert 1\right\rangle \right]  \simeq\cos\left(  \frac{\pi
}{8}\right)  \left\vert 0\right\rangle +e^{i\left(  \frac{2E}{\hslash}%
t-\frac{\pi}{2}\right)  }\sin\left(  \frac{\pi}{8}\right)  \left\vert
1\right\rangle \text{.}%
\end{equation}
Therefore, after some algebraic calculations, we obtain that the spherical
angles $\theta\left(  t\right)  $ and $\varphi\left(  t\right)  $ with
$\left\vert \psi\left(  t\right)  \right\rangle =\left\vert \psi\left(
\theta\left(  t\right)  \text{, }\varphi\left(  t\right)  \right)
\right\rangle $ are given by%
\begin{equation}
\theta\left(  t\right)  =\frac{\pi}{4}\text{, and }\varphi\left(  t\right)
=\frac{2E}{\hslash}t-\frac{\pi}{2}\text{.}%
\end{equation}
It follows then that the accessed and accessible volumes, $\overline
{\mathrm{V}}$ and \textrm{V}$_{\max}$, respectively, are given by%
\begin{align}
\overline{\mathrm{V}}  &  =\frac{2E}{\pi\hslash}\int_{0}^{\frac{\pi\hslash
}{2E}}\left\vert \frac{\sin\left(  \theta\right)  }{2}\left[  \varphi\left(
t\right)  -\varphi\left(  0\right)  \right]  \right\vert dt\nonumber\\
&  =\frac{4E}{\pi\hslash}\int_{0}^{\frac{\pi\hslash}{2E}}\frac{1}{\sqrt{2}%
}\frac{E}{\hslash}tdt\nonumber\\
&  =\frac{\pi}{4\sqrt{2}}%
\end{align}
and,%
\begin{equation}
\mathrm{V}_{\max}=\frac{1}{2}\int_{-\frac{\pi}{2}}^{\frac{\pi}{2}}\sin
(\frac{\pi}{4})d\varphi=\frac{\sqrt{2}\pi}{4}\text{,}%
\end{equation}
since $dV=(1/2)\sin(\theta)d\varphi$ with $\theta=\pi/4$ and $\left(
\theta_{\min}\text{, }\theta_{\max}\text{; }\varphi_{\min}\text{, }%
\varphi_{\max}\right)  =\left(  \pi/4\text{, }\pi/4\text{; }-\pi/2\text{, }%
\pi/2\right)  $. Therefore, the complexity of this evolution reduces to
\textrm{C}$\overset{\text{def}}{=}(\mathrm{V}_{\max}-\overline{\mathrm{V}%
})/\mathrm{V}_{\max}=1/2$. Moreover, given that $s(0$, $\pi\hslash
/(2E))=(2/\hslash)\cdot\Delta E\cdot\left(  \pi\hslash/(2E)\right)  =\pi
/\sqrt{2}$ since $\Delta E=E/\sqrt{2}$, we get from Eq. (\ref{lisa}) that
\textrm{L}$_{\mathrm{C}}=\pi$.

\subsubsection{Connecting $\left\vert C\right\rangle $ and $\left\vert
D\right\rangle $ in Fig. $5$}

In our second part of this second comparative analysis, let us suppose that
the states $\left\vert C\right\rangle $ and $\left\vert D\right\rangle $ are
given by%
\begin{equation}
\left\vert C\right\rangle \overset{\text{def}}{=}\frac{\sqrt{2-\sqrt{2}}}%
{2}\left\vert 0\right\rangle +i\frac{\sqrt{2+\sqrt{2}}}{2}\left\vert
1\right\rangle \text{, and }\left\vert D\right\rangle \overset{\text{def}}%
{=}\frac{\sqrt{2-\sqrt{2}}}{2}\left\vert 0\right\rangle -i\frac{\sqrt
{2+\sqrt{2}}}{2}\left\vert 1\right\rangle \text{,}%
\end{equation}
respectively. The Bloch vectors that correspond to $\left\vert C\right\rangle
$ and $\left\vert D\right\rangle $ are given by $\hat{c}\overset{\text{def}%
}{=}\left(  0\text{, }\sin\left(  3\pi/4\right)  \text{, }\cos\left(
3\pi/4\right)  \right)  =\left(  0\text{, }1/\sqrt{2}\text{, }-1/\sqrt
{2}\right)  $ and $\hat{d}\overset{\text{def}}{=}\left(  0\text{, }\sin\left(
-3\pi/4\right)  \text{, }\cos\left(  -3\pi/4\right)  \right)  =\left(
0\text{, }-1/\sqrt{2}\text{, }-1/\sqrt{2}\right)  $, respectively. From these
expressions of $\hat{c}$ and $\hat{d}$, we find that \textrm{H}$\left(
\alpha\right)  $ is given by%
\begin{equation}
\mathrm{H}\left(  \alpha\right)  \overset{\text{def}}{=}-E\left[  \cos\left(
\alpha\right)  \hat{z}+\sin(\alpha)\hat{x}\right]  \cdot
\mathbf{\boldsymbol{\sigma}}\text{.} \label{h2}%
\end{equation}

\paragraph{Optimal Evolution}

The optimal time evolution from $\left\vert C\right\rangle $ to $\left\vert
D\right\rangle $ occurs under the Hamiltonian \textrm{H}$_{\mathrm{opt}}%
=E\hat{n}_{\mathrm{opt}}\cdot\mathbf{\boldsymbol{\sigma}}$, with $\hat
{n}_{\mathrm{opt}}=\hat{c}\times\hat{d}/\left\Vert \hat{c}\times\hat
{d}\right\Vert =-\hat{x}$. Therefore, \textrm{H}$_{\mathrm{opt}}%
=\mathrm{H}\left(  \pi/2\right)  =-E\sigma_{x}$. After some calculations, it
turns out that the evolution from $\left\vert C\right\rangle $ to $\left\vert
D\right\rangle $ occurs from $t_{i}=0$ to $t_{f}=\pi\hslash/(4E)$ along the
state path on the Bloch sphere specified by the state $\left\vert \psi\left(
t\right)  \right\rangle =e^{\frac{i}{\hslash}E\sigma_{x}t}\left\vert
C\right\rangle $ is given by%
\begin{equation}
\left\vert \psi\left(  t\right)  \right\rangle =\left[  \frac{\sqrt{2-\sqrt
{2}}}{2}\cos\left(  \frac{E}{\hslash}t\right)  -\frac{\sqrt{2+\sqrt{2}}}%
{2}\sin\left(  \frac{E}{\hslash}t\right)  \right]  \left\vert 0\right\rangle
+i\left(  \frac{\sqrt{2-\sqrt{2}}}{2}\sin\left(  \frac{E}{\hslash}t\right)
+\frac{\sqrt{2+\sqrt{2}}}{2}\cos\left(  \frac{E}{\hslash}t\right)  \right)
\left\vert 1\right\rangle \text{.}%
\end{equation}
Therefore, after some algebra, we get that the spherical angles $\theta\left(
t\right)  $ and $\varphi\left(  t\right)  $ with $\left\vert \psi\left(
t\right)  \right\rangle =\left\vert \psi\left(  \theta\left(  t\right)
\text{, }\varphi\left(  t\right)  \right)  \right\rangle $ are given by%
\begin{equation}
\theta\left(  t\right)  =2\arctan\left(  \frac{\frac{\sqrt{2-\sqrt{2}}}{2}%
\sin\left(  \frac{E}{\hslash}t\right)  +\frac{\sqrt{2+\sqrt{2}}}{2}\cos\left(
\frac{E}{\hslash}t\right)  }{\frac{\sqrt{2-\sqrt{2}}}{2}\cos\left(  \frac
{E}{\hslash}t\right)  -\frac{\sqrt{2+\sqrt{2}}}{2}\sin\left(  \frac{E}%
{\hslash}t\right)  }\right)  \text{, and }\varphi=\frac{\pi}{2}\text{,}%
\end{equation}
respectively. Observe that $\theta\left(  0\right)  =3\pi/4$, $\theta\left(
\pi\hslash/(4E)\right)  =-3\pi/4$. Moreover, we have%
\begin{equation}
\lim_{t\rightarrow\frac{\pi\hslash}{8E}^{-}}2\arctan\left(  \frac{\frac
{\sqrt{2-\sqrt{2}}}{2}\sin\left(  \frac{E}{\hslash}t\right)  +\frac
{\sqrt{2+\sqrt{2}}}{2}\cos\left(  \frac{E}{\hslash}t\right)  }{\frac
{\sqrt{2-\sqrt{2}}}{2}\cos\left(  \frac{E}{\hslash}t\right)  -\frac
{\sqrt{2+\sqrt{2}}}{2}\sin\left(  \frac{E}{\hslash}t\right)  }\right)  =\pi
\end{equation}
and,
\begin{equation}
\lim_{t\rightarrow\frac{\pi\hslash}{8E}^{+}}2\arctan\left(  \frac{\frac
{\sqrt{2-\sqrt{2}}}{2}\sin\left(  \frac{E}{\hslash}t\right)  +\frac
{\sqrt{2+\sqrt{2}}}{2}\cos\left(  \frac{E}{\hslash}t\right)  }{\frac
{\sqrt{2-\sqrt{2}}}{2}\cos\left(  \frac{E}{\hslash}t\right)  -\frac
{\sqrt{2+\sqrt{2}}}{2}\sin\left(  \frac{E}{\hslash}t\right)  }\right)
=-\pi\text{.}%
\end{equation}
Therefore, the range of values of the polar angle $\theta$ can be described
as
\begin{equation}
\theta\in\left[  \frac{3}{4}\pi,\pi\right]  _{0\leq t\leq\frac{\pi}{8}}%
\cup\left[  -\pi,-\frac{3}{4}\pi\right]  _{\frac{\pi}{8}\leq t\leq\frac{\pi
}{4}}\text{.}%
\end{equation}
It follows then that the accessed and accessible volumes, $\overline
{\mathrm{V}}$ and \textrm{V}$_{\max}$, respectively, are given by%
\begin{align}
\overline{\mathrm{V}} &  =\frac{8E}{\pi\hslash}\int_{0}^{\frac{\pi\hslash}%
{8E}}\left\vert \frac{\theta\left(  t\right)  -\theta\left(  0\right)  }%
{2}\right\vert dt+\frac{8E}{\pi\hslash}\int_{\frac{\pi\hslash}{8E}}^{\frac
{\pi\hslash}{4E}}\left\vert \frac{\theta\left(  t\right)  -\theta\left(
\frac{\pi\hslash}{8E}\right)  }{2}\right\vert dt\nonumber\\
&  =\frac{8E}{\pi\hslash}\int_{0}^{\frac{\pi\hslash}{8E}}\left\vert
\arctan\left(  \frac{\frac{\sqrt{2-\sqrt{2}}}{2}\sin\left(  \frac{E}{\hslash
}t\right)  +\frac{\sqrt{2+\sqrt{2}}}{2}\cos\left(  \frac{E}{\hslash}t\right)
}{\frac{\sqrt{2-\sqrt{2}}}{2}\cos\left(  \frac{E}{\hslash}t\right)
-\frac{\sqrt{2+\sqrt{2}}}{2}\sin\left(  \frac{E}{\hslash}t\right)  }\right)
-\frac{3}{8}\pi\right\vert dt\nonumber\\
&  +\frac{8E}{\pi\hslash}\int_{\frac{\pi\hslash}{8E}}^{\frac{\pi\hslash}{4E}%
}\left\vert \arctan\left(  \frac{\frac{\sqrt{2-\sqrt{2}}}{2}\sin\left(
\frac{E}{\hslash}t\right)  +\frac{\sqrt{2+\sqrt{2}}}{2}\cos\left(  \frac
{E}{\hslash}t\right)  }{\frac{\sqrt{2-\sqrt{2}}}{2}\cos\left(  \frac
{E}{\hslash}t\right)  -\frac{\sqrt{2+\sqrt{2}}}{2}\sin\left(  \frac{E}%
{\hslash}t\right)  }\right)  +\frac{\pi}{2}\right\vert dt\nonumber\\
&  =\frac{\pi}{8}%
\end{align}
and,%
\begin{align}
\mathrm{V}_{\max} &  =\frac{1}{2}\left[  \left(  \int_{\frac{3}{4}\pi}^{\pi
}d\theta\right)  +\left(  \int_{-\pi}^{-\frac{3}{4}\pi}d\theta\right)
\right]  \nonumber\\
&  =\frac{1}{2}\left(  \frac{\pi}{4}+\frac{\pi}{4}\right)  \nonumber\\
&  =\frac{\pi}{4}\text{,}%
\end{align}
since $dV=(1/2)d\theta$, $\left(  \theta_{\min}\text{, }\theta_{\max}\text{;
}\varphi_{\min}\text{, }\varphi_{\max}\right)  =\left(  3\pi/4\text{, }%
\pi\text{; }\pi/2\text{, }\pi/2\right)  $ for $0\leq t\leq\left(  \pi
\hslash\right)  /(8E)$, and $\left(  \theta_{\min}\text{, }\theta_{\max
}\text{; }\varphi_{\min}\text{, }\varphi_{\max}\right)  =\left(  -\pi\text{,
}-3\pi/4\text{; }\pi/2\text{, }\pi/2\right)  $ for $\left(  \pi\hslash\right)
/(8E)\leq t\leq\left(  \pi\hslash\right)  /(4E)$. Therefore, the complexity of
this evolution reduces to \textrm{C}$\overset{\text{def}}{=}(\mathrm{V}_{\max
}-\overline{\mathrm{V}})/\mathrm{V}_{\max}=1/2$. Moreover, given that $s(0$,
$\pi\hslash/(4E))=2\arccos\left[  \sqrt{(1+\hat{c}\cdot\hat{d})/2}\right]
=\pi/2$, we get from Eq. (\ref{lisa}) that \textrm{L}$_{\mathrm{C}}=\pi
/\sqrt{2}$.

\paragraph{Sub-optimal Evolution}

For a sub-optimal time evolution from $\left\vert C\right\rangle $ to
$\left\vert D\right\rangle $, we consider the Hamiltonian \textrm{H}%
$_{\text{\textrm{sub}-}\mathrm{opt}}=\mathrm{H}\left(  0\right)  =-E\sigma
_{z}$. After some calculations, it turns out that the evolution from
$\left\vert C\right\rangle $ to $\left\vert D\right\rangle $ occurs from
$t_{i}=0$ to $t_{f}=\pi\hslash/(2E)$ along the state path on the Bloch sphere
specified by the state $\left\vert \psi\left(  t\right)  \right\rangle
=e^{\frac{i}{\hslash}E\sigma_{z}t}\left\vert C\right\rangle $ given by%
\begin{equation}
\left\vert \psi\left(  t\right)  \right\rangle =e^{i\frac{E}{\hslash}t}%
\frac{\sqrt{2-\sqrt{2}}}{2}\left\vert 0\right\rangle +e^{-i\frac{E}{\hslash}%
t}\frac{\sqrt{2+\sqrt{2}}}{2}\left\vert 1\right\rangle \simeq\frac
{\sqrt{2-\sqrt{2}}}{2}\left\vert 0\right\rangle +\frac{\sqrt{2+\sqrt{2}}}%
{2}e^{-i\frac{2E}{\hslash}t}\left\vert 1\right\rangle \text{.}%
\end{equation}
Therefore, after some algebra, we get that the spherical angles $\theta\left(
t\right)  $ and $\varphi\left(  t\right)  $ with $\left\vert \psi\left(
t\right)  \right\rangle =\left\vert \psi\left(  \theta\left(  t\right)
\text{, }\varphi\left(  t\right)  \right)  \right\rangle $ are given by%
\begin{equation}
\theta\left(  t\right)  =\frac{3\pi}{4}\text{, and }\varphi\left(  t\right)
=\frac{2E}{\hslash}t\text{,}%
\end{equation}
respectively. It follows then that the accessed and accessible volumes,
$\overline{\mathrm{V}}$ and \textrm{V}$_{\max}$, respectively, are given by%
\begin{align}
\overline{\mathrm{V}}  &  =\frac{2E}{\pi\hslash}\int_{0}^{\frac{\pi\hslash
}{2E}}\left\vert \frac{\sin\left(  \frac{3\pi}{4}\right)  }{2}\left[
\varphi\left(  t\right)  -\varphi\left(  0\right)  \right]  \right\vert
dt\nonumber\\
&  =\frac{2E}{\pi\hslash}\int_{0}^{\frac{\pi\hslash}{2E}}\frac{1}{\sqrt{2}%
}\frac{E}{\hslash}tdt\nonumber\\
&  =\frac{\pi}{4\sqrt{2}}%
\end{align}
and,%
\begin{equation}
\mathrm{V}_{\max}=\frac{1}{2}\int_{0}^{\pi}\sin(\frac{3\pi}{4})d\varphi
=\frac{\pi}{2\sqrt{2}}\text{,}%
\end{equation}
since $dV=(1/2)\sin(\theta)d\varphi$ with $\theta=(3/4)\pi$ and $\left(
\theta_{\min}\text{, }\theta_{\max}\text{; }\varphi_{\min}\text{, }%
\varphi_{\max}\right)  =\left(  3\pi/4\text{, }3\pi/4\text{; }0\text{, }%
\pi\right)  $. Therefore, the complexity of this evolution reduces to
\textrm{C}$\overset{\text{def}}{=}(\mathrm{V}_{\max}-\overline{\mathrm{V}%
})/\mathrm{V}_{\max}=1/2$. Furthermore, given that $s(0$, $\pi\hslash
/(2E))=(2/\hslash)\cdot\Delta E\cdot\left(  \pi\hslash/(2E)\right)  =\pi
/\sqrt{2}$ since $\Delta E=E/\sqrt{2}$, we get from Eq. (\ref{lisa}) that
\textrm{L}$_{\mathrm{C}}=\pi$.

In Table II, we illustrate the numerical values of the geodesic efficiency
$\eta_{\mathrm{GE}}$ in Eq. (\ref{jap}), the speed efficiency $\eta
_{\mathrm{SE}}$ in Eq. (\ref{se2}), the curvature coefficient $\kappa
_{\mathrm{AC}}^{2}$ in Eq. (\ref{XXXX}), the complexity \textrm{C }in Eq.
(\ref{QCD}) and, lastly, the complexity length scale \textrm{L}$_{\mathrm{C}}$
in Eq. (\ref{complexityL}) for the two pairs of quantum evolutions considered
here, with each pair being specified by an optimal time and a sub-optimal time
quantum evolution.\begin{table}[t]
\centering
\begin{tabular}
[c]{c|c|c|c|c|c|c|c|c}\hline\hline
\textbf{Hamiltonian} & \textbf{Initial state} & \textbf{Final state} &
\textbf{Travel time} & $\eta_{\mathrm{GE}}$ & $\eta_{\mathrm{SE}}$ &
$\kappa_{\mathrm{AC}}^{2}$ & \textrm{C} & \textrm{L}$_{\mathrm{C}}$\\\hline
$-E\sigma_{x}$, time optimal & $\frac{\sqrt{2+\sqrt{2}}}{2}\left\vert
0\right\rangle -i\frac{\sqrt{2-\sqrt{2}}}{2}\left\vert 1\right\rangle $ &
$\frac{\sqrt{2+\sqrt{2}}}{2}\left\vert 0\right\rangle +i\frac{\sqrt{2-\sqrt
{2}}}{2}\left\vert 1\right\rangle $ & $\frac{\pi\hslash}{4E}$ & $1$ & $1$ &
$0$ & $\frac{1}{2}$ & $\frac{\pi}{\sqrt{2}}$\\\hline
$E\sigma_{z}$, time sub-optimal & $\frac{\sqrt{2+\sqrt{2}}}{2}\left\vert
0\right\rangle -i\frac{\sqrt{2-\sqrt{2}}}{2}\left\vert 1\right\rangle $ &
$\frac{\sqrt{2+\sqrt{2}}}{2}\left\vert 0\right\rangle +i\frac{\sqrt{2-\sqrt
{2}}}{2}\left\vert 1\right\rangle $ & $\frac{\pi\hslash}{2E}$ & $\frac
{1}{\sqrt{2}}$ & $\frac{1}{\sqrt{2}}$ & $4$ & $\frac{1}{2}$ & $\pi$\\\hline
$-E\sigma_{x}$, time optimal & $\frac{\sqrt{2-\sqrt{2}}}{2}\left\vert
0\right\rangle +i\frac{\sqrt{2+\sqrt{2}}}{2}\left\vert 1\right\rangle $ &
$\frac{\sqrt{2-\sqrt{2}}}{2}\left\vert 0\right\rangle -i\frac{\sqrt{2+\sqrt
{2}}}{2}\left\vert 1\right\rangle $ & $\frac{\pi\hslash}{4E}$ & $1$ & $1$ &
$0$ & $\frac{1}{2}$ & $\frac{\pi}{\sqrt{2}}$\\\hline
$-E\sigma_{z}$, time sub-optimal & $\frac{\sqrt{2-\sqrt{2}}}{2}\left\vert
0\right\rangle +i\frac{\sqrt{2+\sqrt{2}}}{2}\left\vert 1\right\rangle $ &
$\frac{\sqrt{2-\sqrt{2}}}{2}\left\vert 0\right\rangle -i\frac{\sqrt{2+\sqrt
{2}}}{2}\left\vert 1\right\rangle $ & $\frac{\pi\hslash}{2E}$ & $\frac
{1}{\sqrt{2}}$ & $\frac{1}{\sqrt{2}}$ & $4$ & $\frac{1}{2}$ & $\pi$\\\hline
\end{tabular}
\caption{\emph{Comparing optimal and sub--optimal time evolutions}. Numerical
values of the geodesic efficiency $\eta_{\mathrm{GE}}$, the speed efficiency
$\eta_{\mathrm{SE}}$, the curvature coefficient $\kappa_{\mathrm{AC}}^{2}$,
the complexity \textrm{C} and, lastly, the complexity length scale
\textrm{L}$_{\mathrm{C}}$ for four quantum evolutions. The first two
evolutions, for instance, connect the same (distinct) initial and final
states. However, while the first evolution is time optimal, the second one is
not. The same comparative description applies to the pair specified \ by the
third and fourth quantum evolutions that appear in our tabular summary.
Finally, as expected on physical grounds, the first (second) and third
(fourth) quantum evolutions perform identically in terms of the measures
$\eta_{\mathrm{GE}}$, $\eta_{\mathrm{SE}}$, $\kappa_{\mathrm{AC}}^{2}$,
\textrm{C}, and \textrm{L}$_{\mathrm{C}}$.}%
\end{table}We are now ready for our conclusions.

\section{Final Remarks}

We present here a summary of our findings along with remarks on their
relevance and limitations. Finally, we propose future research directions
originating from our investigation.

\subsection{Summary of results}

In this paper, we advanced our quantitative understanding of the intricacies
involved in both time-optimal and time sub-optimal quantum Hamiltonian
evolutions that link arbitrary source and target states on the Bloch sphere,
as recently discussed in Ref. \cite{carloNPB}. To begin, we selected and
analyzed unitary Schr\"{o}dinger quantum evolutions through various metrics,
including path length, geodesic efficiency, speed efficiency, and the
curvature coefficient of the corresponding quantum-mechanical trajectory
connecting the source state to the target state on the Bloch sphere. Following
this, we assessed the chosen evolutions using our proposed complexity measure,
as well as in relation to the concept of complexity length scale. The
selection of both time-optimal and time sub-optimal evolutions, along with the
choice of source and target states, allowed us to perform relevant sanity
checks aimed at validating the physical significance of the framework
underpinning our proposed complexity measure. Our findings indicate that,
generally, efficient quantum evolutions exhibit lower complexity compared to
their inefficient counterparts. Nonetheless, it is crucial to acknowledge that
complexity is not exclusively dictated by length; indeed, longer trajectories
that are sufficiently curved may demonstrate a complexity that is less than or
equal to that of shorter trajectories with a lower curvature coefficient.

\medskip

In this paper, we have significantly enhanced our comprehension of the
importance\textbf{ }of symmetry and invariance arguments concerning a
complexity measure from a physical standpoint. Indeed, for a given class of
initial and final states together with suitably chosen stationary Hamiltonian
evolutions on the Bloch sphere, this comprehension was achieved in two
quantitative steps. In the first step (Fig. $4$ and Table I), we verified that
the complexity $\mathrm{C}(\left\vert A\right\rangle \overset{\mathrm{H}%
_{\mathrm{opt}}}{\longrightarrow}\left\vert B\right\rangle )$ of the evolution
from an initial state $\left\vert A\right\rangle $ to a final state
$\left\vert B\right\rangle $ under the time optimal Hamiltonian $\mathrm{H}%
_{\mathrm{opt}}$ is equal to the complexity $\mathrm{C}(\left\vert A^{\prime
}\right\rangle \overset{\mathrm{H}_{\mathrm{opt}}^{\prime}}{\longrightarrow
}\left\vert B^{\prime}\right\rangle )$ of the evolution from the source state
$\left\vert A^{\prime}\right\rangle $ to the target state $\left\vert
B^{\prime}\right\rangle $ under the time optimal Hamiltonian $\mathrm{H}%
_{\mathrm{opt}}^{\prime}$ provided that the distance $s_{0}^{AB}=s_{\left\vert
A\right\rangle \rightarrow\left\vert B\right\rangle }^{\mathrm{opt}}$ along
the shortest geodesic path that joins $\left\vert A\right\rangle $ to
$\left\vert B\right\rangle $ equals the distance $s_{0}^{A^{\prime}B^{\prime}%
}=s_{\left\vert A^{\prime}\right\rangle \rightarrow\left\vert B^{\prime
}\right\rangle }^{\mathrm{opt}}$ along the shortest geodesic path that joins
$\left\vert A^{\prime}\right\rangle $ to $\left\vert B^{\prime}\right\rangle
$. Stated otherwise, we checked that $\mathrm{C}(\left\vert A\right\rangle
\overset{\mathrm{H}_{\mathrm{opt}}}{\longrightarrow}\left\vert B\right\rangle
)=\mathrm{C}(\left\vert A^{\prime}\right\rangle \overset{\mathrm{H}%
_{\mathrm{opt}}^{\prime}}{\longrightarrow}\left\vert B^{\prime}\right\rangle
)$, if $s_{0}^{AB}=s_{0}^{A^{\prime}B^{\prime}}$. In the second step (Fig. $5$
and Table II), instead, we checked that the complexity $\mathrm{C}(\left\vert
A\right\rangle \overset{\mathrm{H}_{\mathrm{sub}\text{-}\mathrm{opt}}%
}{\longrightarrow}\left\vert B\right\rangle )$ of the evolution from
$\left\vert A\right\rangle $ to $\left\vert B\right\rangle $ under the time
optimal Hamiltonian $\mathrm{H}_{\mathrm{sub}\text{-}\mathrm{opt}}$ is equal
to the complexity $\mathrm{C}(\left\vert A^{\prime}\right\rangle
\overset{\mathrm{H}_{\mathrm{sub}\text{-}\mathrm{opt}}^{\prime}}%
{\longrightarrow}\left\vert B^{\prime}\right\rangle )$ of the evolution from
$\left\vert A^{\prime}\right\rangle $ to $\left\vert B^{\prime}\right\rangle $
under the time optimal Hamiltonian $\mathrm{H}_{\mathrm{sub}\text{-}%
\mathrm{opt}}^{\prime}$ provided that the distance $s^{AB}=s_{\left\vert
A\right\rangle \rightarrow\left\vert B\right\rangle }^{\mathrm{sub}%
\text{-}\mathrm{opt}}$ along the effective dynamical trajectory on the Bloch
sphere that joins $\left\vert A\right\rangle $ to $\left\vert B\right\rangle $
equals the distance $s^{A^{\prime}B^{\prime}}=s_{\left\vert A^{\prime
}\right\rangle \rightarrow\left\vert B^{\prime}\right\rangle }^{\mathrm{sub}%
\text{-}\mathrm{opt}}$ along the effective dynamical trajectory that joins
$\left\vert A^{\prime}\right\rangle $ to $\left\vert B^{\prime}\right\rangle
$. In other words, we checked that $\mathrm{C}(\left\vert A\right\rangle
\overset{\mathrm{H}_{\mathrm{sub}\text{-}\mathrm{opt}}}{\longrightarrow
}\left\vert B\right\rangle )=\mathrm{C}(\left\vert A^{\prime}\right\rangle
\overset{\mathrm{H}_{\mathrm{sub}\text{-}\mathrm{opt}}^{\prime}}%
{\longrightarrow}\left\vert B^{\prime}\right\rangle )$, if $s^{AB}%
=s^{A^{\prime}B^{\prime}}$ and $s_{0}^{AB}=s_{0}^{A^{\prime}B^{\prime}}$, with
$s_{0}^{AB}\leq s^{AB}$ and $s_{0}^{A^{\prime}B^{\prime}}\leq s^{A^{\prime
}B^{\prime}}$. Furthermore, our work shows that although the concepts of
length and complexity of a quantum evolution between two given states on the
Bloch sphere are distinct quantities, there is an intrinsic link between them
dictated by fundamental symmetry and invariance arguments of physical origin.

\subsection{Limitations and outlook}

Our study is subject to two notable limitations. Firstly, we have confined our
analysis to quantum evolutions that are governed by stationary Hamiltonians
within the framework of two-level quantum systems. Secondly, we have not
addressed the implications of our findings for the characterization of
complexity in quantum-mechanical systems that exist in mixed quantum states.
While extending our research to time-varying Hamiltonian evolutions is
theoretically feasible, the computational challenge lies in deriving precise
analytical solutions to the time-dependent Schr\"{o}dinger equation for
calculating probability amplitudes. This task presents considerable
difficulties, even within the realm of two-level quantum systems
\cite{landau32,zener32,rabi37,rabi54,barnes12,barnes13,messina14,grimaudo18,cafaroijqi,castanos,grimaudo23,elena20}%
. Additionally, transitioning from pure to mixed states introduces various
challenges, both computationally and conceptually. It is well recognized that
there is an infinite array of distinguishability metrics applicable to mixed
quantum states \cite{karol06}. This diversity leads to interpretations of
critical geometric quantities, such as the complexity and volume of quantum
states, which are contingent upon the selected metric. Specifically, the
non-uniqueness of these distinguishability measures necessitates a thorough
understanding of the physical implications associated with the choice of a
particular metric, a topic that holds significant conceptual and practical
relevance \cite{silva21,mera22,luongo24,chien24}. Given that these
complexities exceed the scope of our current investigation, we aim to address
certain aspects of these challenges in our future research endeavors.

\medskip

The current study, while focused on the complexity analysis of geometric
styles limited to single-qubit evolutions as demonstrated in Ref.
\cite{marrone19}, opens avenues for numerous additional investigations.
Initially, our comparison was restricted to geodesic unwasteful evolutions
versus nongeodesic wasteful evolutions. However, the range of dynamical
options is significantly broader \cite{campa19,campaioli19,xu24}. Notably, it
would be valuable to examine (in terms of complexity metrics) geodesic
unwasteful paths against geodesic wasteful paths, or alternatively, against
nongeodesic unwasteful paths. Furthermore, we can broaden our comparative
analysis to include nonstationary Hamiltonian evolutions characterized by
relevant time-dependent magnetic field configurations
\cite{messina14,grimaudo18,cafaroijqi,castanos,grimaudo23}. Additionally, we
stress that our analysis is confined to a two-state system, which currently
prevents us from exploring the relationship between our complexity measure and
quantum entanglement. Nevertheless, as an initial step towards this goal, we
are in the process of broadening our methodology to encompass\textbf{ }%
$d$\textbf{-}level quantum systems where $d$\textbf{ }exceeds\textbf{ }%
$2$\textbf{, }within larger finite-dimensional Hilbert spaces
\cite{jakob01,kimura03,krammer08,kurzy11,xie20,siewert21}. Finally, we may
investigate the potential for a comparative analysis of the complexity of
quantum evolutions within the Bloch sphere for quantum systems exhibiting
mixed quantum states \cite{hornedal22,nade24}.

\medskip

While we have introduced our own method for quantifying the complexity of
quantum evolutions, it is important to acknowledge the established
methodologies present in the literature, as mentioned in the Introduction.
Notably, two prominent approaches are those based on Krylov's and Nielsen's
complexities. Despite their effectiveness, a comprehensive understanding of
the similarities and differences between these two metrics remains an area for
further investigation. Recent findings indicate that, although these
complexities cannot be equated \cite{rolph24}, they can be represented using a
common matrix structure \cite{craps24}. Specifically, the all-time average of
Krylov's complexity related to state evolution and the late-time upper bound
of Nielsen's complexity for the evolution operator can both be articulated
through a particular matrix formalism \cite{craps24}. However, it is crucial
to note that Krylov and Nielsen complexities cannot be considered equivalent;
Nielsen complexity functions as a distance measure, whereas Krylov complexity
does not adhere to the triangle inequality and is not a distance measure
between states or operators \cite{rolph24}. Nonetheless, recent evidence has
demonstrated that the square root of Krylov complexity serves as a distance
measure between single qubit states linked by unitary time evolution and
satisfies the triangle inequality \cite{nath24}. In particular, the square
root of Krylov's complexity is proportional to the trace distance, which is
also proportional to the standard Euclidean distance between single qubit
states on the Bloch sphere. Before presenting our conclusive remark, we would
like to highlight that substantial research regarding the (Krylov) complexity
of integrable and chaotic quantum Hamiltonian evolutions is available in Refs.
\cite{bala20a,bala21a,rabinovici22,craps22a,craps24b,craps25}. Krylov
complexity is an important metric for evaluating how quickly quantum operators
spread across the full range of possible operators during dynamic evolution.
It is expected that the late-time plateau of this metric will help distinguish
between integrable and chaotic dynamics; however, its ability to do so is
highly dependent on the choice of the initial seed. In Ref. \cite{craps25},
for example, the authors propose an enhancement to Krylov complexity that
employs not just a single operator as the starting point, but rather a
comprehensive collection of all simple operators within the theoretical
framework. These operators are selected based on a well-defined physical
criterion, such as the inclusion of all few-body operators. The late-time
plateau of this modified measure consistently shows lower values for
integrable systems in comparison to chaotic ones, as demonstrated through a
series of standard test cases typically used in the analysis of quantum chaos.
We believe that the rationale for incorporating our sanity checks in this
study presented here bears a resemblance to the motivations behind the
multiseed Krylov complexity method discussed in Ref. \cite{craps25}.
Currently, this is regarded as a belief, and a complete understanding of this
relationship requires an in-depth examination, which we will postpone for
future studies.

\bigskip

In summary, considering that our complexity measure is derived from volumes
and the fascinating opportunity to examine a generalization of the triangle
inequality in relation to volumes \cite{rolph24}, it would be beneficial to
investigate this type of inequality within our framework to enhance our
understanding of the proposed complexity measure from a physical perspective.

\textbf{\medskip}

Our research, despite its limitations, has the potential to motivate other
scientists to explore this area further. We are confident that our findings
may pave the way for new extensions and applications in the future. A more
comprehensive discussion regarding these possible advancements will be
reserved for future scientific endeavors.

\begin{acknowledgments}
The authors wish to extend their sincere appreciation to the anonymous
referees for their valuable feedback, which has significantly improved the
quality of the paper. C.C. expresses his gratitude to Paul M. Alsing and
Newshaw Bahreyni for valuable discussions. Additionally,\textbf{ }the authors
express their gratitude to Leonardo Rossetti for his technical support in the
Python implementation of Figures 3, 4, and 5. Lastly, E.C. acknowledges the
UAlbany Summer Undergraduate Research Program for providing her with a
Fellowship in 2024, during which she engaged with several topics addressed in
this paper. Any opinions, findings and conclusions or recommendations
expressed in this material are those of the author(s) and do not necessarily
reflect the views of their home Institutions.
\end{acknowledgments}

\end{document}